\documentclass[preprintnumbers,amsmath,amssymb,showpacs]{revtex4}
\usepackage{epsfig}
\begin{document}
\setlength{\voffset}{1.0cm}
\title{Breathers and their interaction in the massless Gross-Neveu model}
\author{Christian Fitzner\footnote{christian.fitzner@gravity.fau.de}}
\author{Michael Thies\footnote{michael.thies@gravity.fau.de}}
\affiliation{Institut f\"ur Theoretische Physik III,
Universit\"at Erlangen-N\"urnberg, D-91058 Erlangen, Germany}
\date{\today}
\begin{abstract}
The breather is a vibrating multifermion bound state of the massless Gross-Neveu model, originally found by Dashen, Hasslacher and Neveu
in the large $N$ limit. We exhibit the salient features of this state and confirm that it solves the relativistic time-dependent Hartree-Fock
equations. We then solve the scattering problem of two breathers with arbitrary internal parameters and velocities, generalizing an ansatz
recently developed for the baryon-baryon scattering problem in the same model. The exact analytical solution is given and illustrated with
a few examples.
\end{abstract}
\maketitle

\section{Introduction}\label{sect1}
The principal sources of experimental information about strong interactions are hadron spectroscopy and hadronic scattering processes
at accelerators. Theoretically, spectroscopy is by now fairly well understood owing to lattice simulations of quantum chromodynamics,
at least for the low-lying, most stable states. Since it is necessary to work in Euclidean time when computing the path integral by Monte-Carlo
methods, this tool fails in the case of scattering problems. As a consequence, there has been little progress towards understanding 
scattering of composite, relativistic bound states in the non-perturbative regime over the last decades. Early attempts to use hadron 
models of confined quarks for scattering were plagued by difficulties with covariance. It is expected that only a relativistic quantum field 
theory can account correctly for covariance. Then one is immediately faced with the lack of calculational tools at strong coupling in 
Minkowski space. This is one of the motivations for us to step back and study scattering processes of composite objects in an exactly 
solvable toy model, the Gross-Neveu (GN) model \cite{L1}.

As is well known, the GN model is the 1+1 dimensional relativistic quantum field theory of $N$ flavors of massless Dirac fermions, interacting through 
a scalar-scalar contact interaction, 
\begin{equation}
{\cal L} =   \sum_{k=1}^N \bar{\psi}_k i\partial \!\!\!/ \psi_k + \frac{g^2}{2}\left( \sum_{k=1}^N \bar{\psi}_k \psi_k \right)^2.
\label{1.1}
\end{equation}
A number of exact, analytical results about the static properties of this model have been obtained notably in
the 't~Hooft limit ($N \to \infty,\  Ng^2=$ const.), ranging from the rich hadron spectrum and hadron structure \cite{L2,L3} to the phase
diagram at finite temperature and chemical potential \cite{L4}. During the past few years, time-dependent issues like boosted
hadrons and structure functions \cite{L5} or scattering processes involving kinks \cite{L6,L7}, kink-antikink baryons \cite{L8}
and composite multibaryon states \cite{L9} have been solved exactly. Although a full mathematical proof of the
recent results in \cite{L9} is still missing, it is probably fair to say that we understand the scattering of
any number of boosted, static hadrons of arbitrary complexity in the initial and final states.

As pointed out in \cite{L9}, this cannot be the whole story, even in such a simple toy model as the GN model. Dashen, Hasslacher and Neveu
(DHN) have already discovered a ``breather" solution long ago \cite{L2}, i.e., a multifermion bound state oscillating in time in its rest frame.
It generalizes the ground state baryon to a collective excited state. Breathers are a well-studied soliton species in the field of nonlinear
science, the best known example probably being the sine-Gordon breather, a vibrating kink-antikink state. The theoretical interest in this
particular kind of solitons stems from the role they play in many different areas, such as condensed matter physics, hydrodynamics 
and nonlinear optics, see e.g. \cite{L10,L11,L12,L13,L14} and references therein. (One particularly nice application is to view the breather 
as a moving, relativistic clock for the
twin paradox of special relativity \cite{L15}.) Breathers are less familiar in particle physics, as they have no obvious correspondence 
in the known particle zoo. The reason lies in their genuine classical character. In the GN model, they owe their existence to the large $N$ limit,
an idealization which is quite far from reality in particle physics. A useful analogy in strong interaction physics can nevertheless be identified,
namely the collective vibration of a heavy nucleus. In the large $N$ limit, hadrons may be thought of as systems made of 
a large number of constituents which can then exhibit classical behavior, much like heavy nuclei or molecules in the real world. 

The main difference between the DHN breather and the sine-Gordon breather is the fact that the GN model is a fermionic theory. Here,
the breather describes the mean field (Hartree-Fock potential), generated dynamically by quantized fermions that it drags along. 
We cannot start from a classical, bosonic, non-linear equation like the sine-Gordon equation, but have to solve a quantum mechanical 
self-consistency 
problem, including the polarization of the Dirac sea. Describing how one can find such solutions systematically and clarifying the role
of the fermions populating the breather, also in a breather-breather scattering process, will be the main topics of this paper. 

In Sect.~\ref{sect2}, we will review in detail the DHN breather. This particular solution of the GN model has never been discussed in any detail, 
to the best of our knowledge. In Ref. \cite{L6}, it has only served to derive kink-antikink scattering by analytic continuation
in one parameter, following a suggestion of DHN in their original work \cite{L2}. The main part of the paper, Sect.~\ref{sect3}, 
is then devoted to the intricate scattering process of two breathers. This covers at the same time breather-breather bound states
as well as all problems that can be arrived at by replacing one or both of the breathers by a DHN baryon or a kink.  This whole class of problems
will be solved exactly with a generalization of an ansatz method developed for baryon-baryon scattering in \cite{L8}. We finish
with a brief summary and an outlook, Sect.~\ref{sect4}.

\section{Single DHN breather}\label{sect2}

Before attacking the complicated breather-breather scattering problem, we have to get familiar with the breather of the GN model originally 
found by DHN \cite{L2}. The present section serves to introduce the breather, illustrate its salient features and show how to generalize the 
ansatz technique of Ref.~\cite{L8} to breather-type solutions. We also compute the breather
mass and its fermion density using the Hartree-Fock approach. From a technical point of view, we shall use this section to set up a convenient notation, a prerequisite
for the more involved problems to follow.

\subsection{Reminder of the self-consistent scalar potential}\label{sect2a}

In their seminal paper where baryons of the GN model were first constructed \cite{L2}, DHN also report on a time-dependent, 
semi-classical solution, the breather. Whereas the baryons were derived by means of inverse scattering theory, the form of the 
breather self-consistent scalar potential has been guessed by the authors, using the analogy with the well-known sine-Gordon breather. The 
parameters were then determined self-consistently. DHN also suggested that analytic continuation in one of the parameters to 
imaginary values should describe kink-antikink scattering, a suggestion which was taken up and verified in detail in Ref.~\cite{L6}.
There, one can also find expressions for the breather spinors corrected for misprints in the original reference.  In the present 
subsection, we recall the form of the scalar potential and fermion density in the original notation of DHN and illustrate it with a few 
representative examples.

Following previous works, we identify DHN's semi-classical path integral method with the relativistic version of time-dependent 
Hartree-Fock (TDHF) in the canonical framework, i.e., the solution of the self-consistency problem
\begin{equation}
\left( i \partial \!\!\!/ - S\right) \psi_{\alpha} =0, \qquad S=-g^2 \sum_{\beta}^{\rm occ} \bar{\psi}_{\beta}\psi_{\beta}.
\label{2.1}
\end{equation}
The sum runs over all occupied single fermion states, including the Dirac sea. Units in which the vacuum fermion mass is 1 will be used
throughout this work. DHN write the scalar potential as
\begin{equation}
S = \frac{\cosh K x - a \cos \Omega t + b(1-K^2/2)}{\cosh Kx + a \cos \Omega t + b}.
\label{2.2}
\end{equation}
There are two independent variables, $b$ and $\epsilon$. The parameters $\Omega,K$ can be expressed through $\epsilon$ as
\begin{equation}
\Omega  =  \frac{2}{\sqrt{1+\epsilon^2}}, \quad K \ = \ \epsilon \Omega,
\label{2.3}
\end{equation}
whereas $a$ is the solution of the equation
\begin{equation}
0 = b^2 K^4 + 4K^2 \left(1-b^2 \right) + 4\Omega^2 a^2.
\label{2.4}
\end{equation}
We may choose the positive square root without loss of generality, since the other sign merely corresponds to a shift of $t$ by half a period. 
Thus $S$ is fully specified by the parameters $\epsilon>0$ and $b>\sqrt{1+\epsilon^2}$, where the latter bound follows from (\ref{2.4}).
Just like the DHN baryon, the breather has two bound states. 
Since the potential is periodic in time, they are not eigenstates of the Hamiltonian. In analogy to quasi-momenta and the Bloch theorem, we
can define quasi-energies $\omega$ via the Floquet theorem,
\begin{equation}
\psi(t+T) = e^{-i \omega T} \psi(t), \quad T=\frac{2\pi}{\Omega}.
\label{2.4a}
\end{equation}
DHN find a pair of bound states with quasi-energies $\omega = \pm (1+\epsilon^2)^{-1/2}$, reflecting the 
charge conjugation symmetry of the GN model. The lower one is taken to be fully occupied, whereas the upper one carries 
$\nu N$ fermions ($\nu=0...1$ will be referred to as occupation fraction).
The antibreather would have the upper state empty, the lower state occupied by $(1-\nu)N$ fermions, and identical $S$.
We do not consider more general ways of filling the bound states which also exist, just like for baryons. 
DHN find the following self-consistency condition relating $\nu, b$ and $\epsilon$, 
\begin{equation}
b=(1-\nu) \frac{\sqrt{1+\epsilon^2}}{1-(2/\pi)\arctan \epsilon}.
\label{2.5}
\end{equation}
It is worth mentioning that the breather contains the DHN baryon as a special case. For $a=0$ or, equivalently, $b=\sqrt{1+\epsilon^2}$
(its minimal value), $S$ becomes static,
\begin{equation}
S=1 - \frac{2 \epsilon^2}{\sqrt{1+\epsilon^2} \left( \cosh \frac{2 \epsilon x}{\sqrt{1+\epsilon^2}} + \sqrt{1+\epsilon^2} \right)}.
\label{2.6}
\end{equation}
Comparison with the original parametrization of the DHN baryon \cite{L2},
\begin{equation}
S=1 + y (\tanh \xi_- - \tanh \xi_+), \qquad \xi_{\pm} = y x \pm \frac{1}{2} {\rm artanh} y,
\label{2.7}
\end{equation}
shows perfect agreement for the choice
\begin{equation}
y= \frac{\epsilon}{\sqrt{1+\epsilon^2}}.
\label{2.8}
\end{equation}
The quasi-energies of the breather bound states then  go over into the energies of the baryon bound states, $\pm \sqrt{1-y^2}$.
Likewise, the self-consistency condition (\ref{2.5}) reduces to 
\begin{equation}
1-\nu_{\rm stat} = 1- \frac{2}{\pi} \arctan \epsilon,
\label{2.9}
\end{equation}
in agreement with the baryon case. Nothing like this could happen in the case of the sine-Gordon breather. Here, there
is simply no static kink-antikink bound state to which the breather could possibly be reduced. This difference is due to 
the valence fermions in the GN case which overcome the kink-antikink repulsion and lead to bound, static kink-antikink states.

\begin{figure}[h]
\begin{center}
\epsfig{file=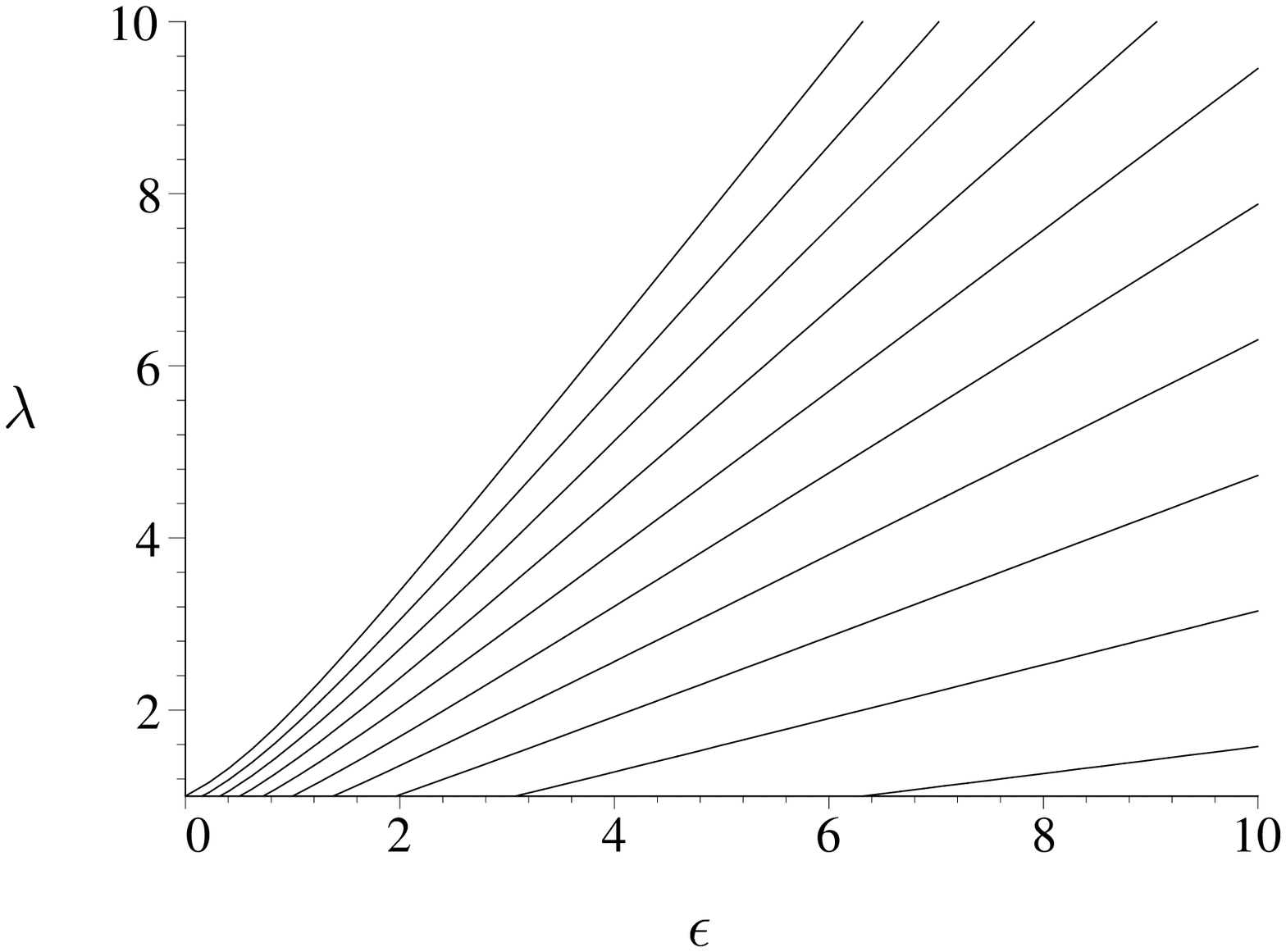,angle=270,width=8cm}
\caption{Allowed breather parameter region in the ($\epsilon,\lambda$) plane and curves of constant occupation fraction $\nu$ 
(from top to bottom: $\nu=n/10, n=0,...,9$). The $\epsilon$ axis ($\lambda=1$) corresponds to the static baryon where all $\nu \in [0,1]$ are
allowed.}
\label{Fig1}
\end{center}
\end{figure}

The physical meaning of the parameters $\epsilon,b$ can be exhibited as follows. Choose a value of $\epsilon>0$. For the minimal
allowed value of $b$ ($b=\sqrt{1+\epsilon^2}$), the breather becomes static and $\epsilon$ determines the size and shape of this baryon
 as well as its occupation.
If we now increase $b$, the frequency of the breather does not change (being solely determined by $\epsilon$), but the amplitude
of the oscillation increases. At the same time, the occupation fraction $\nu$ decreases,
\begin{equation}
\frac{1-\nu}{1-\nu_{\rm stat}} = \frac{b}{\sqrt{1+\epsilon^2}} := \lambda \ge 1.
\label{2.10}
\end{equation}
Let us try to give an overview of how the breather behaves in space and time. 
First consider the allowed range of parameters, using ($\epsilon, \lambda$) rather than ($\epsilon,b$). As illustrated in Fig.~\ref{Fig1},
the physical region in the ($\epsilon,\lambda$) plane is restricted to
\begin{equation}
1 \le \lambda \le \left( 1- \frac{2}{\pi} \arctan \epsilon \right)^{-1}.
\label{2.11}
\end{equation}
These limits are shown together with curves of constant $\nu$ in Fig.~\ref{Fig1}. Along
the upper boundary, fermion number vanishes ($\nu=0$). The lower boundary ($\lambda=1$) corresponds to the baryon where $\epsilon$
and $\nu$ are related according to Eq.~(\ref{2.9}). Thus for a given occupation fraction $\nu$, a one-parameter family of breathers 
exists. By contrast, the sine-Gordon breather has only a single parameter governing both its frequency and amplitude and carries no
fermions.

We illustrate the dynamics of the breather for two values of $\epsilon$. The shape of the baryon 
potential evolves from a shallow, attractive  well  for small $\epsilon$ to a widely separated kink-antikink pair at large $\epsilon$.
The breather potential at $t=0$ is qualitatively similar and oscillates monotonically (but anharmonically) between two limiting curves,
with a period of $T=2\pi/\Omega$. The fermion density of the breather is computed by adding up the contributions from the Dirac sea and the two bound states.
If the lower bound state is completely filled, its contribution is cancelled by the density induced in the Dirac sea and the full density is
simply given by the upper bound state, just like for the baryon. The total fermion number is $N \nu$. 

\begin{figure}[h]
\begin{center}
\epsfig{file=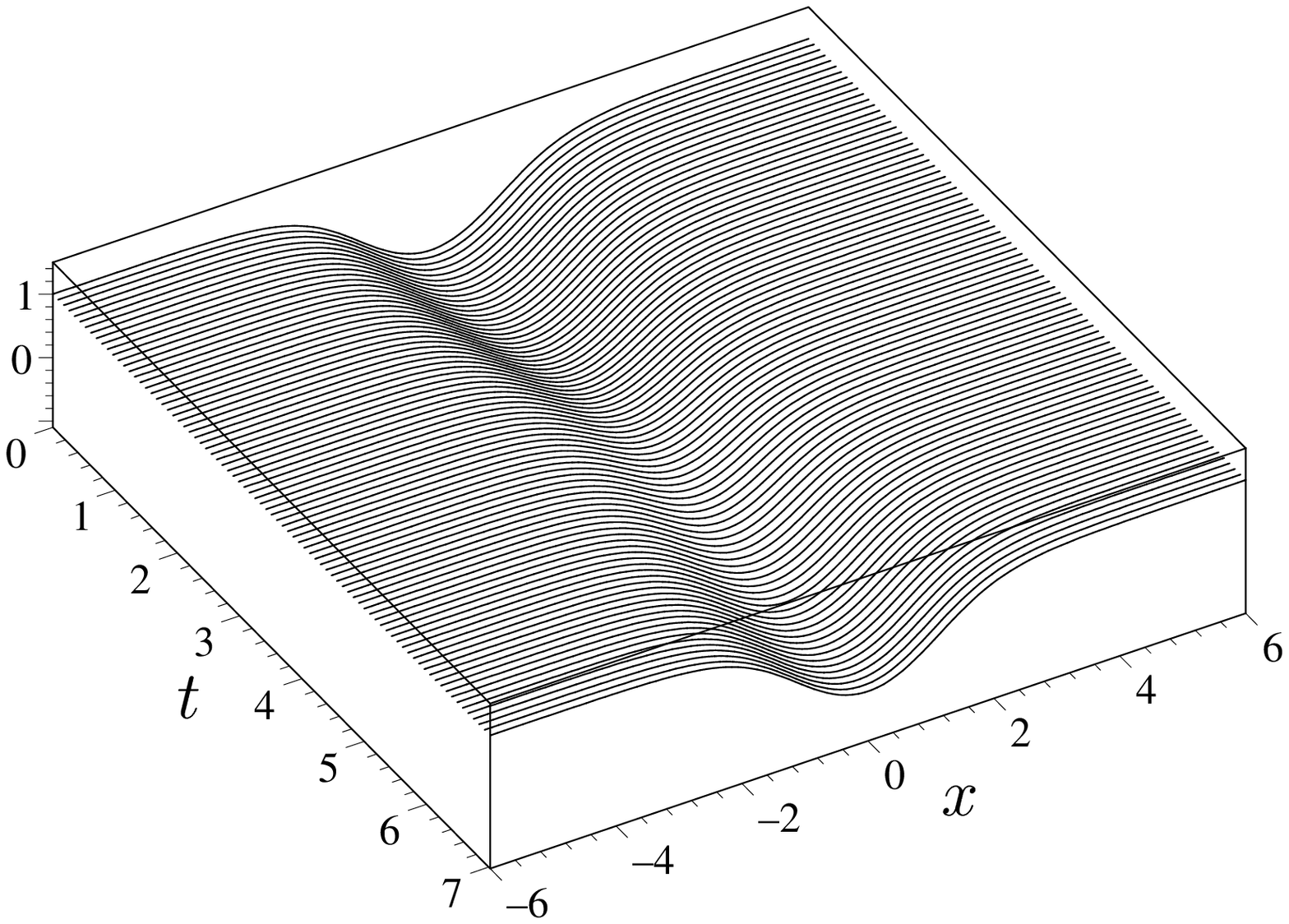,angle=270,width=8cm}\epsfig{file=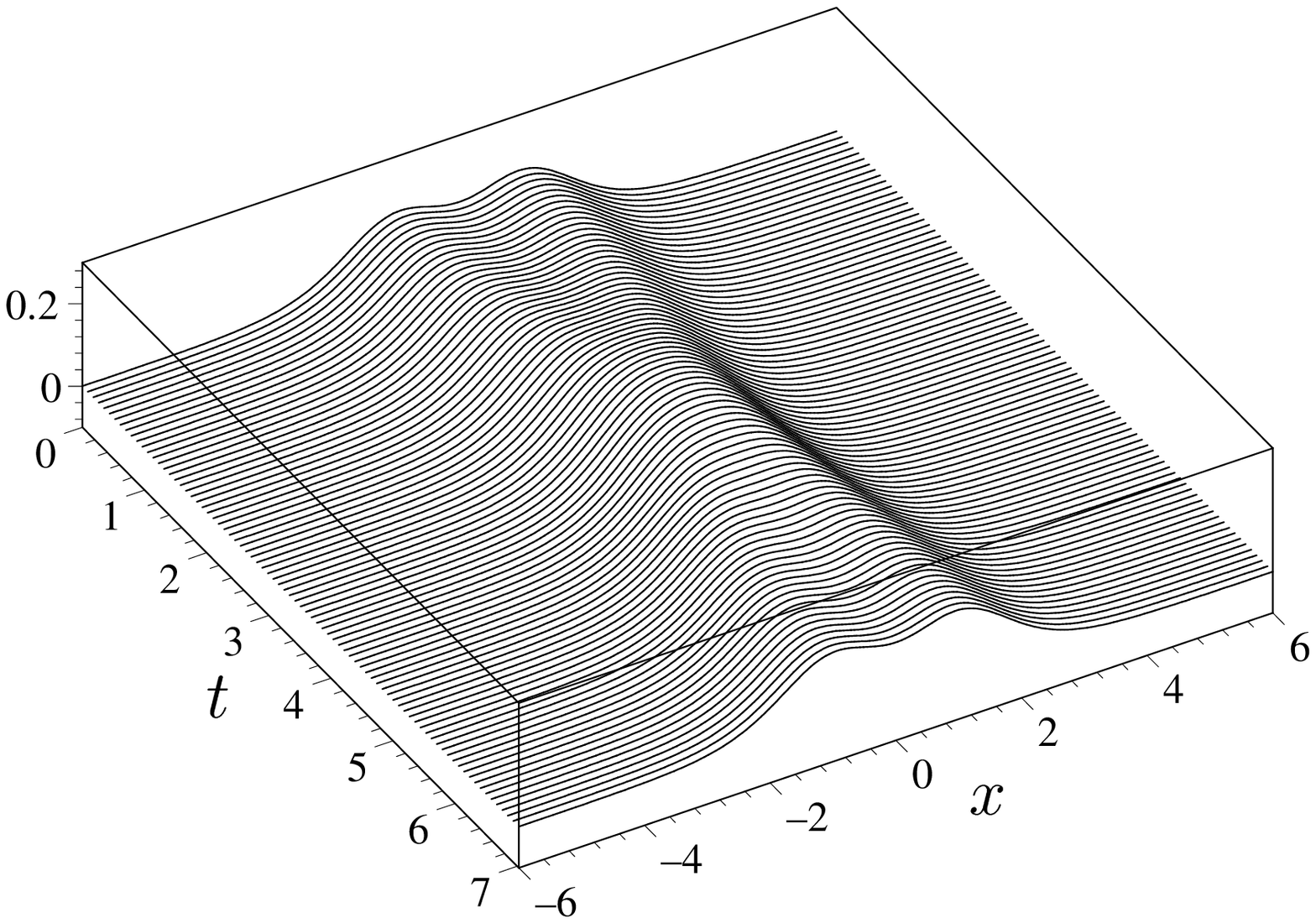,angle=270,width=8cm}
\caption{Scalar potential $S$ (left plot) and fermion density $\rho$ (right plot) during one period of the breather. Parameters: $\epsilon=2, \lambda=1.1$
corresponding to period $T=7.0248$ and occupation fraction $\nu=0.6753$. See \cite{anim} for animations.}
\label{Fig2}
\end{center}
\end{figure}

Let us first illustrate the scalar potential and the fermion density for the moderate value $\epsilon=2$ and two different values of $\lambda$.
For $\lambda=1.1, \nu=0.6753$, $S$ oscillates, staying always below the vacuum value $S=1$, see Fig.~\ref{Fig2}. 
\begin{figure}[h]
\begin{center}
\epsfig{file=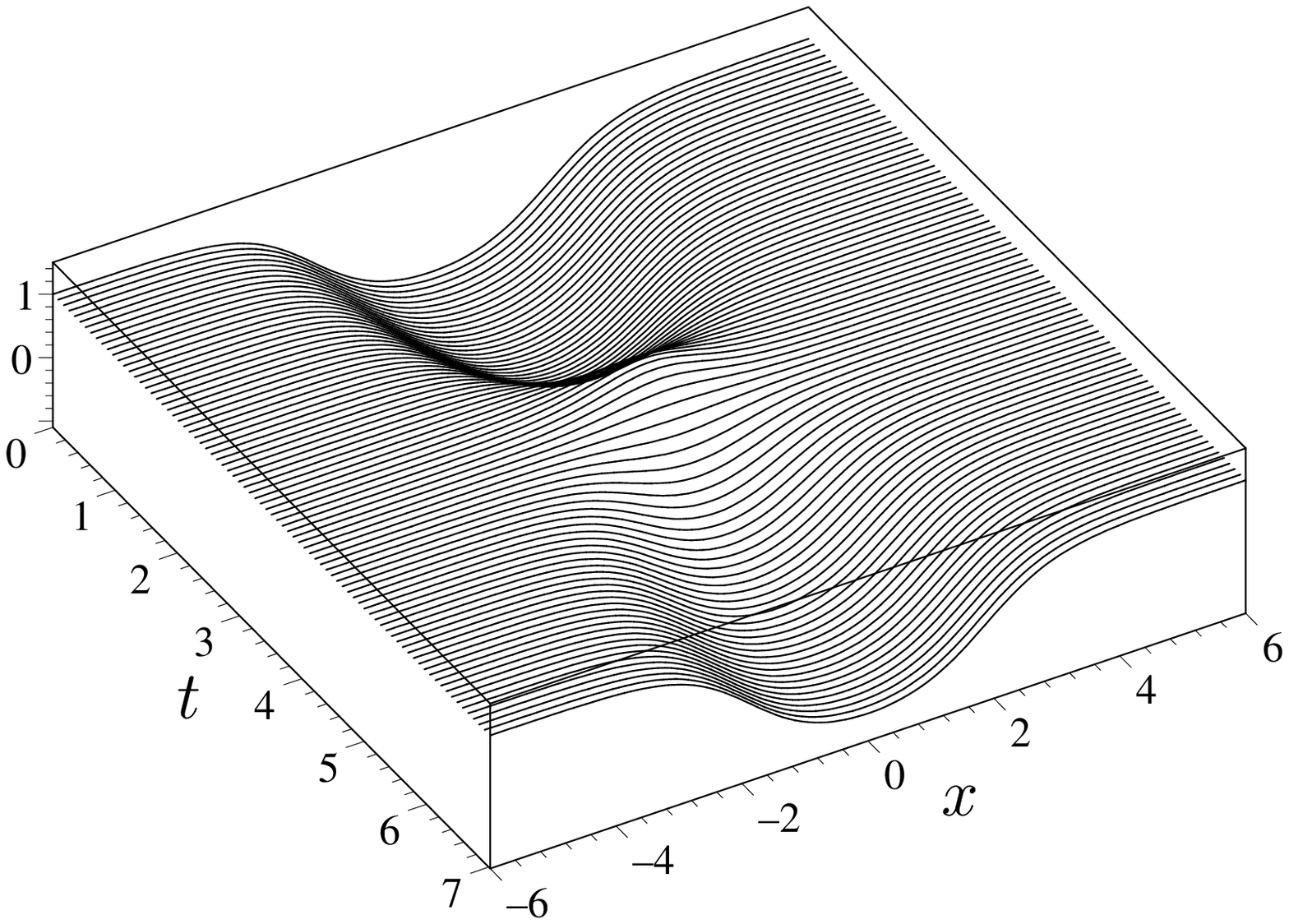,angle=270,width=8cm}\epsfig{file=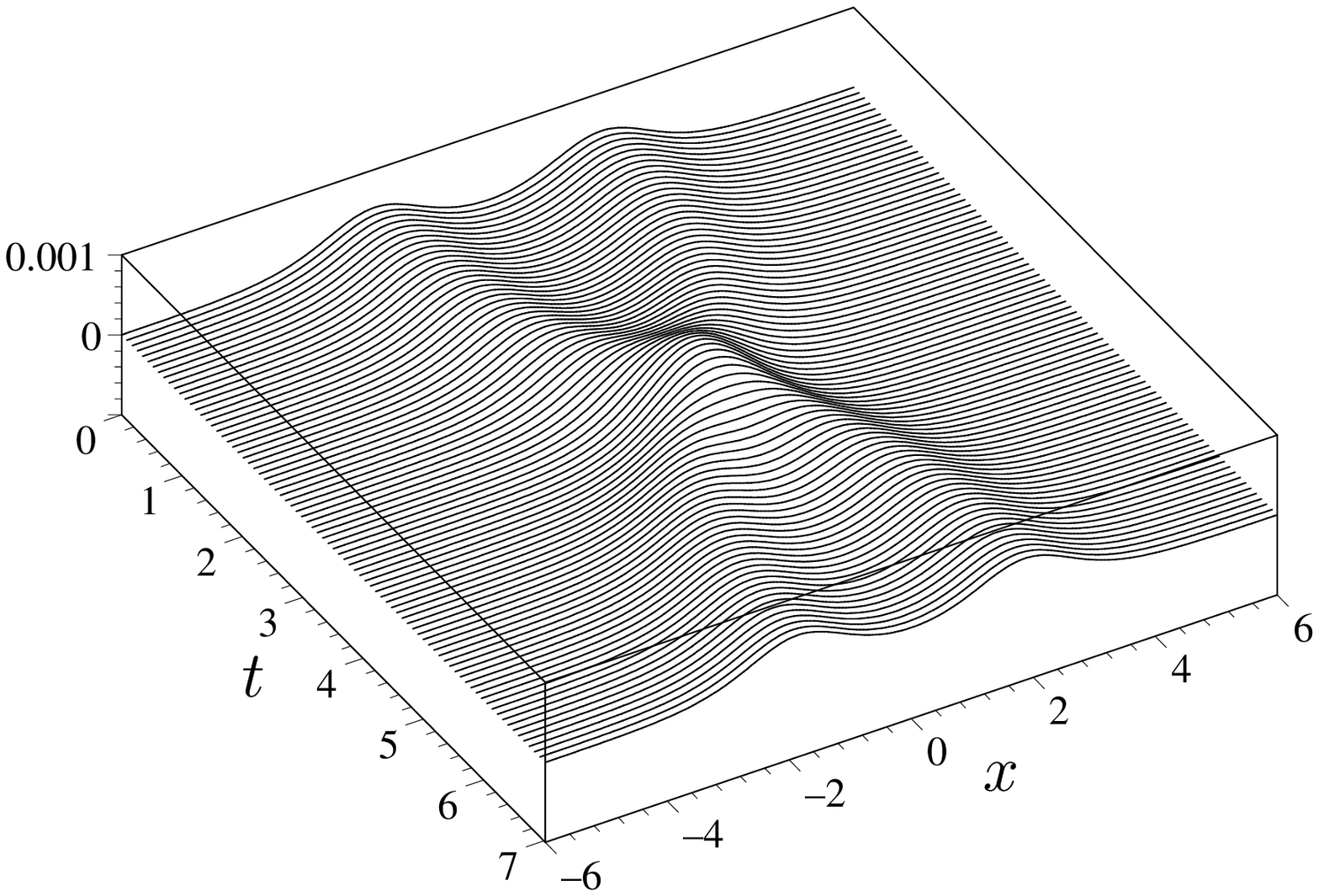,angle=270,width=8cm}
\caption{Like Fig.~\ref{Fig2}, but for $\lambda= 3.38$ corresponding to $\nu=0.002335$. The potential $S$ exceeds the value $1$ near the midpoint by 40$\%$.
Note the different scales in the density plots of Figs.~\ref{Fig2} and \ref{Fig3}. See \cite{anim} for animations.} 
\label{Fig3}
\end{center}
\end{figure}
There, one also sees that the density oscillates between two peaks and a single peak.
If we increase the value of the breather amplitude by choosing $\lambda=3.38, \nu=0.002335$ and the same 
$\epsilon$,  $S$ overshoots near the center of the breather, reaching the value 1.39, see Fig.~\ref{Fig3}. Between the two examples shown in Figs.~\ref{Fig2} and \ref{Fig3}
there is a ``critical value" of $\lambda$, $b=1+\epsilon^2$ or $\lambda=\sqrt{1+\epsilon^2}$, where the potential oscillates
between a baryon-like shape and the constant vacuum value ($S=1$). For this particular value of $b$, $S$ becomes
\begin{equation}
S_{\rm crit} = \frac{\cosh Kx - \epsilon^2 \cos \Omega t + 1 - \epsilon^2}{\cosh Kx + \epsilon^2 \cos \Omega t + 1+ \epsilon^2}.
\label{2.12}
\end{equation}
Indeed, after 1/2 period, $\cos \Omega t= -1$ and $S_{\rm crit}=1$ for all $x$. This special case is noteworthy for yet another reason, namely its
close relationship to the sine-Gordon breather. The sine-Gordon breather is a solution of the sine-Gordon equation
\begin{equation}
\quad \partial_{\mu}\partial^{\mu} \phi + \sin \phi = 0
\label{2.13}
\end{equation}
given by
\begin{equation}
\phi = 4 \arctan \left( \frac{\sqrt{1-\omega^2}\cos \omega t}{\omega \cosh \sqrt{1-\omega^2}x}\right).
\label{2.14}
\end{equation}
One can check that 
\begin{equation}
S_{\rm crit} = \cos \frac{\phi}{2},
\label{2.15}
\end{equation}
provided one identifies $\omega$ with $\Omega/2= 1/\sqrt{1+ \epsilon^2}$. 
The motivation behind the particular nonlinear transformation (\ref{2.15}) is the fact that it also maps the sine-Gordon kink,
$\phi=4 \arctan e^x$, onto the Gross-Neveu antikink, $S=-\tanh x$. We do not know whether this is a mere coincidence or whether 
there is a deeper reason behind this mapping. 

\begin{figure}[h]
\begin{center}
\epsfig{file=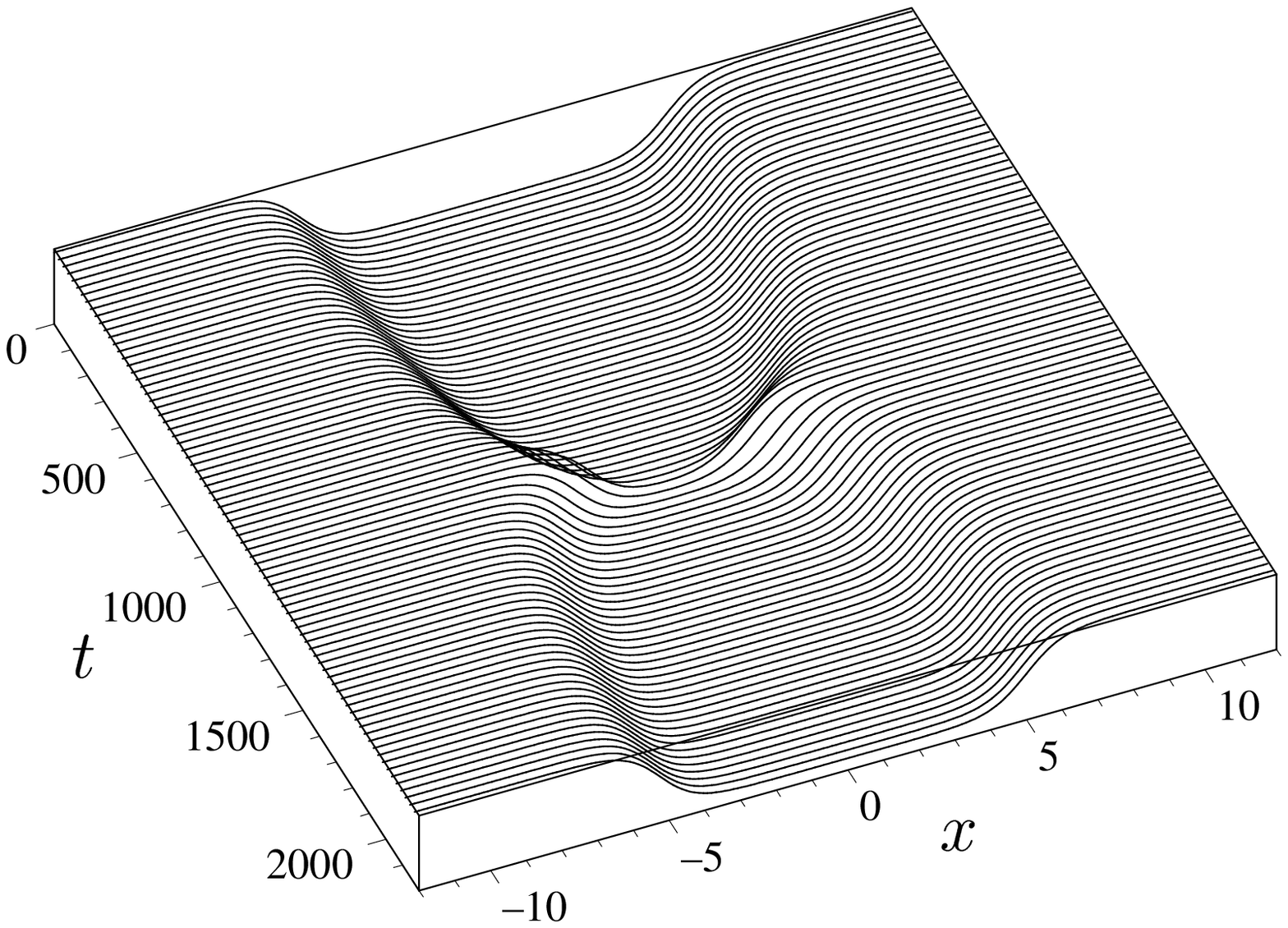,angle=270,width=8cm}\epsfig{file=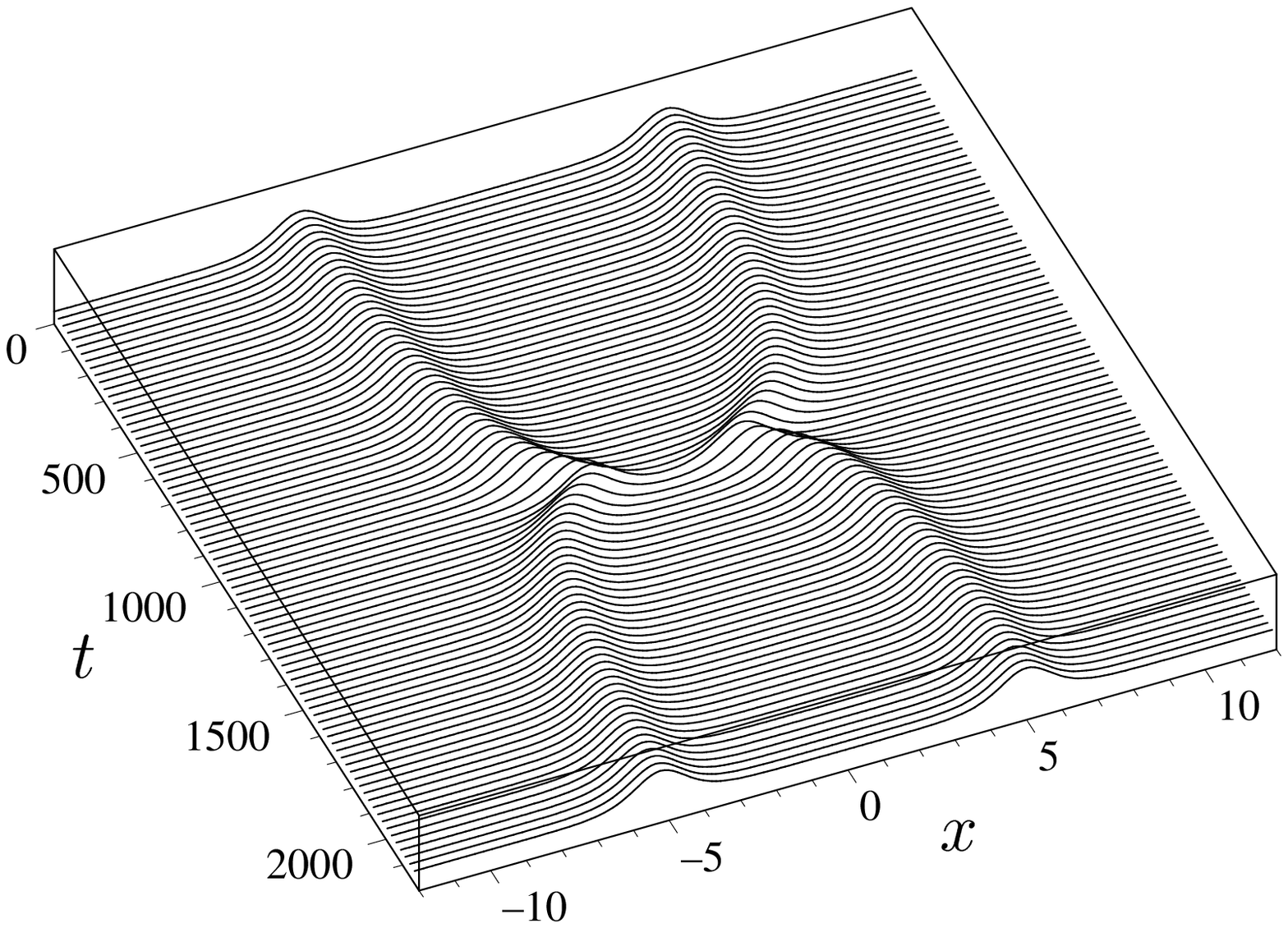,angle=270,width=8cm}
\caption{Scalar potential $S$ (left plot) and fermion density $\rho$ (right plot) during one period of the breather. Parameters: $\epsilon=700, \lambda=10$
corresponding to period $T=2199$ and occupation fraction $\nu=0.9909$. Kink and antikink stay well separated during the whole period. 
See \cite{anim} for animations.}
\label{Fig4}
\end{center}
\end{figure}

\begin{figure}[h]
\begin{center}
\epsfig{file=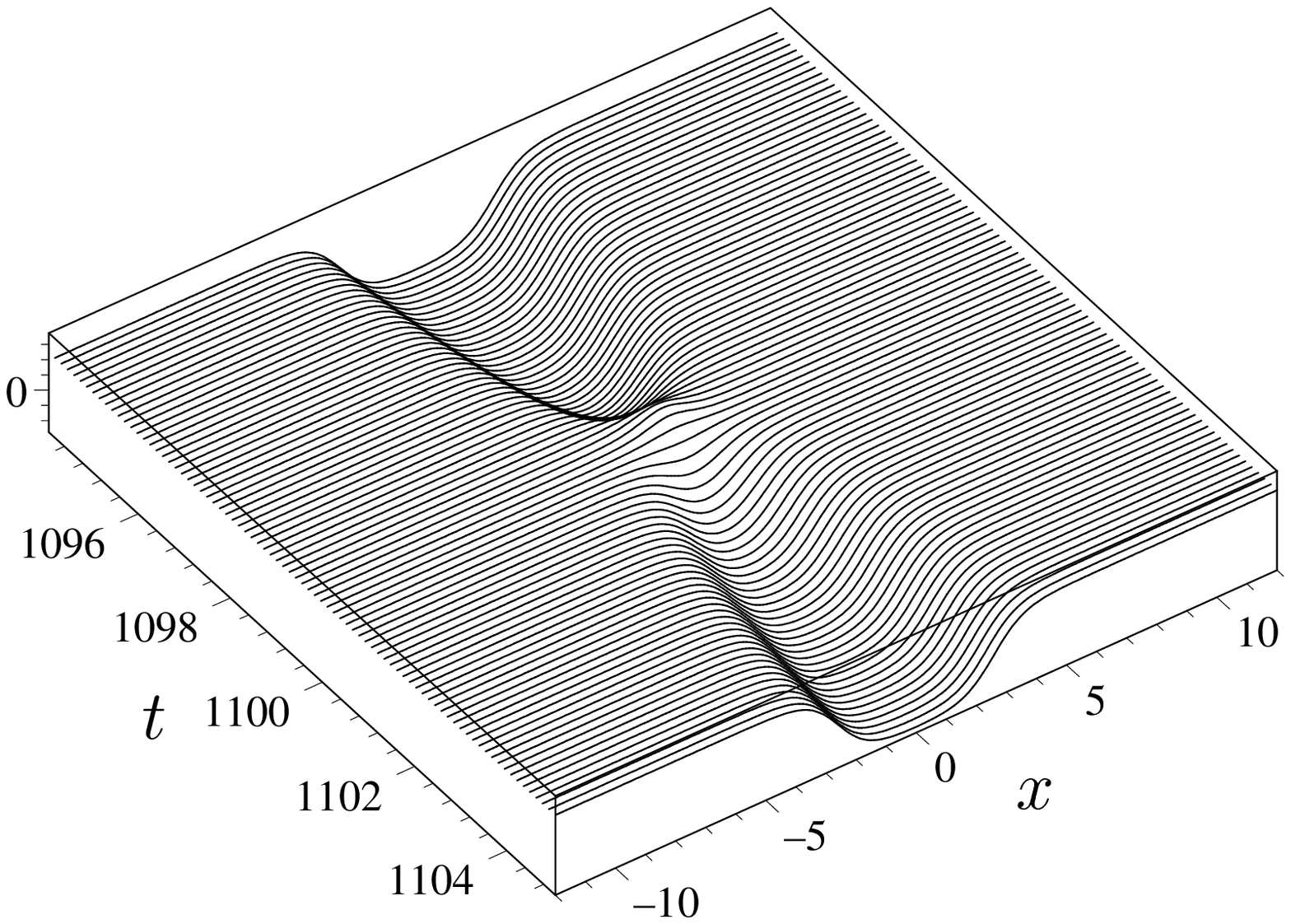,angle=270,width=8cm}\epsfig{file=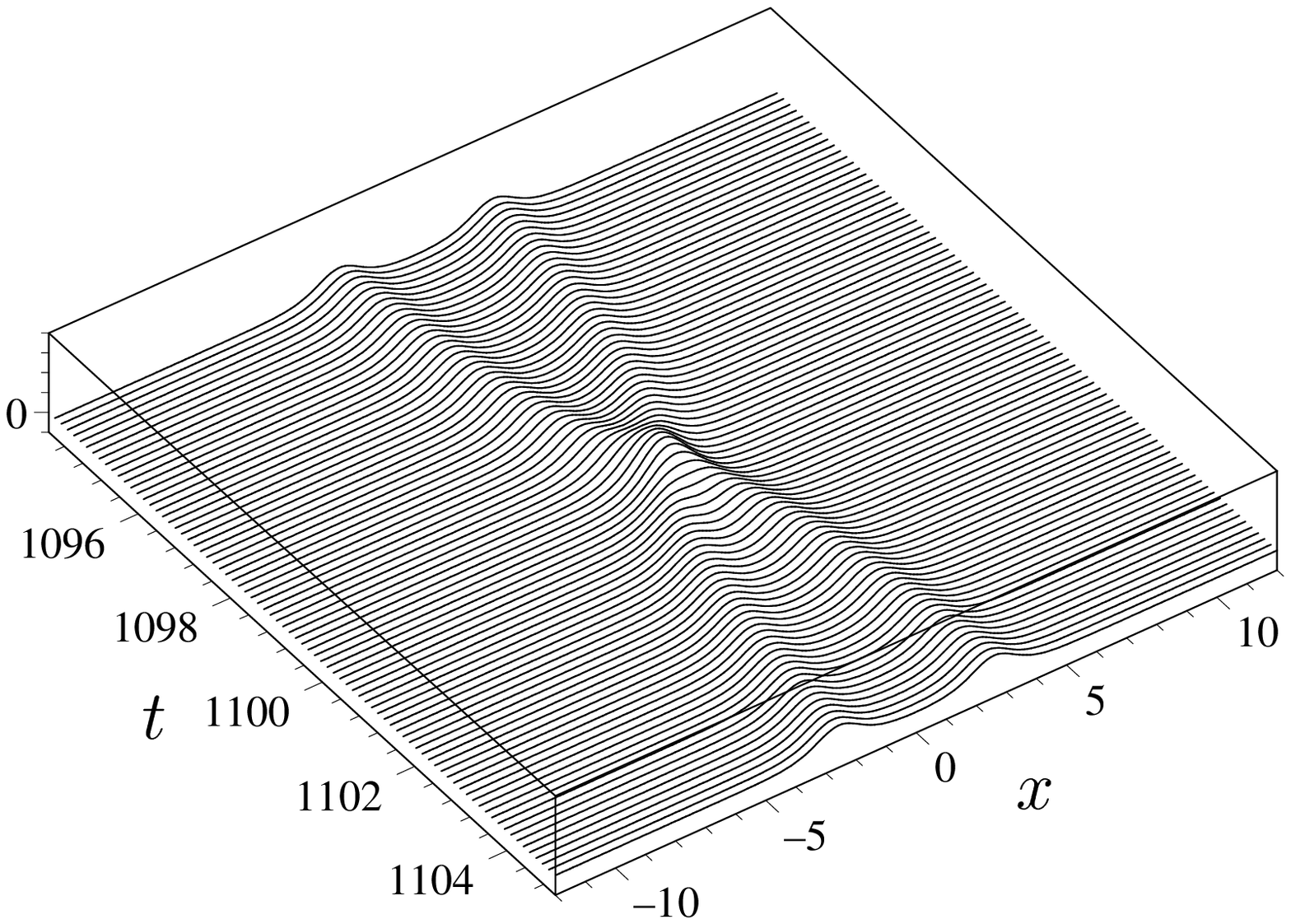,angle=270,width=8cm}
\caption{Like Fig.~\ref{Fig4}, but for $\lambda=1000, \nu=0.09054$. The overshooting takes place in a tiny fraction of the full period ($T=2199$)  
shown here, indicating strong anharmonicity of the breather. See \cite{anim} for animations.}
\label{Fig5}
\end{center}
\end{figure}

We now turn to a large value of $\epsilon$ ($\epsilon=700$) where the breather has a more pronounced kink-antikink shape.
Fig.~\ref{Fig4} shows plots of $S$ and the fermion density at $\lambda=10, \nu=0.9909$, where the system exhibits oscillation of a kink against
an antikink without overshooting. 
If we go beyond the critical value $\lambda=\sqrt{1+\epsilon^2}$, $S$ again exceeds the value of 1 at the midpoint.
However the time scale for this overshooting is now much shorter than the period of the breather, indicating that the motion is strongly
anharmonic. For this reason it is difficult to present a plot similar to Fig.~\ref{Fig3} here. Instead, we show the very short time
interval where $S$ exceeds 1 in ``slow motion",  cf. Fig.~\ref{Fig5}. 

\subsection{Light cone variables and rational parameters}\label{sect2b}

The DHN breather has 2 real parameters, $\epsilon$ and $b$, governing its size, frequency and amplitude in
the rest frame. In addition, one can perform arbitrary Poincar\'e transformations, yielding 3 additional parameters (velocity, shift
in $x$ and $t$). Breather-breather scattering then will depend on 10 parameters, as compared to 6 parameters for
baryon-baryon scattering. In view of this complication and according to our prior experience with baryon scattering \cite{L8,L9}, it is crucial 
to simplify the kinematics as much as possible by going to light cone variables. Moreover, it is highly advisable to introduce parameters
such that the final results are rational functions, rather than algebraic or transcendental functions. This is a prerequisite for being able
to reproduce lengthy computer algebra (CA) computations independently and simplify them to a unique form. In this 
subsection, we will take over the notation from previous papers whenever possible and extend the rational parametrizations from the baryon
 to the breather.

Our convention for light cone variables is
\begin{equation}
z=x-t, \quad \bar{z}=x+t, \quad \partial_0 = \bar{\partial}-\partial, \quad \partial_1 = \bar{\partial}+\partial.
\label{2.16}
\end{equation}
We use the light cone spectral parameter
\begin{equation}
k = \frac{1}{2} \left( \zeta- \frac{1}{\zeta} \right), \quad \omega = - \frac{1}{2}  \left( \zeta+ \frac{1}{\zeta} \right),
\label{2.17}
\end{equation}
rather than momentum $k$ and energy $\omega$.
The boost parameter $\eta$, velocity $v$ and rapidity $\xi$ are related by
\begin{equation}
\eta = e^{\xi} = \sqrt{\frac{1+v}{1-v}}, \quad v=\frac{\eta^2-1}{\eta^2+1}.
\label{2.18}
\end{equation}
Under a boost, the light cone coordinates and spectral parameter are simply rescaled,
\begin{equation}
z \to \eta z, \quad \bar{z}\to \eta^{-1} \bar{z} ,\quad  \zeta \to \eta \zeta.
\label{2.19}
\end{equation}
In these variables the Lorentz invariant argument of a plane wave reads
\begin{equation}
kx-\omega t = \frac{1}{2} \left( \zeta \bar{z} - \frac{z}{\zeta} \right).
\label{2.20}
\end{equation}
Together with the chiral choice of Dirac matrices (diagonal $\gamma_5$),
\begin{equation}
\gamma^0 = \sigma_1, \quad \gamma^1 = i \sigma_2, \quad \gamma_5 = \gamma^0 \gamma^1 = - \sigma_3,
\label{2.21}
\end{equation}
the Dirac equation entering the TDHF approach simplifies to
\begin{equation}
2 i \bar{\partial} \psi_2 = S \psi_1, \quad 2i \partial \psi_1 = - S \psi_2,
\label{2.22}
\end{equation}
with chirality $\psi_1 = \psi_L, \psi_2 = \psi_R$. 
In the case of the DHN baryon, the following reparametrization of $y$ has proven to be useful to avoid the appearance of square 
roots,
\begin{equation}
y_1 = \frac{Z_1^2-1}{2iZ_1}, \quad w_1 = \sqrt{1-y_1^2} = - \frac{Z_1^2+1}{2Z_1}, \quad Z_1\in {\rm U}(1).
\label{2.23}
\end{equation}
Introducing the phase $\varphi_1$,
\begin{equation}
Z_1 = e^{i\varphi_1}, \quad y_1= \sin \varphi_1, \quad w_1 = - \cos \varphi_1,
\label{2.24}
\end{equation}
we see that one should restrict $Z_1$ to the 2nd quadrant ($\pi/2 \le \varphi_1 \le \pi$) in order to parametrize the relevant
region $y_1,w_1\ge 0$. These are all the ingredients without which the solution of baryon scattering problems would have
been almost hopelessly complicated. In the same vein, let us reparametrize the breather specific quantities
($K, \Omega, \epsilon$) and ($a, b, \lambda$). For $K,\Omega,\epsilon$, we choose again a phase $Z\in {\rm U}(1)$ via
\begin{equation}
K= \frac{Z^2-1}{iZ}, \quad \Omega = - \frac{Z^2+1}{Z}, \quad \epsilon = \frac{K}{\Omega} = i \left( \frac{Z^2-1}{Z^2+1} \right).
\label{2.25}
\end{equation}
Upon setting $Z=e^{i\varphi}$, we find
\begin{equation}
K=2\sin \varphi, \quad \Omega= - 2 \cos \varphi, \quad \epsilon = - \tan \varphi, \quad \sqrt{1+\epsilon^2} = - \frac{1}{\cos \varphi},
\label{2.26}
\end{equation}
so that $K,\Omega,\epsilon>0$ corresponds to $\varphi \in [\pi/2, \pi]$. The remaining set of parameters ($a,b,\lambda$) require yet another
phase variable $Q \in {\rm U}(1)$ in the 2nd quadrant. We set
\begin{equation}
a= - \frac{Z^2-1}{Z^2+1}\frac{Q^2-1}{Q^2+1}, \quad b=\frac{2Z}{Z^2+1}\frac{2Q}{Q^2+1}
\label{2.27}
\end{equation}
so that Eq.~(\ref{2.4}) is satisfied identically. Parametrizing $Q=e^{i\psi}$, one finds
\begin{equation}
a=\tan \varphi \tan \psi, \quad b = \frac{1}{\cos \varphi \cos \psi}, \quad \lambda=- \frac{1}{\cos \psi}, \quad
\sqrt{\lambda^2-1} = - \tan \psi,
\label{2.28}
\end{equation}
hence $a,b,\lambda>0$ is consistent with $\psi \in [\pi/2,\pi]$. Thus the two new parameters ($Z,Q$) replacing ($\epsilon, b$)
both live on the quarter of the unit circle in the 2nd quadrant.
The baryon limit of the breather ($b=\sqrt{1+\epsilon^2}$) amounts to setting $Q=-1$. 
In this limit, $Z$ has the same meaning as in our previous works on baryon scattering \cite{L8,L9}. 
The value $b=1+\epsilon^2$ where the DHN breather can be mapped onto the sine-Gordon breather corresponds to the choice
$Q=Z$. 

\subsection{Breather mass}\label{sect2c}

The mass of the breather has been computed by DHN \cite{L2}. We find it instructive to repeat the calculation in the Hartree-Fock approach, closely following Ref.~\cite{L5} for
the baryon mass. As explained there, it is advantageous to compute the total energy by 
splitting the energy density into local (vacuum subtracted) and constant pieces. The local part (which gives a finite contribution when
integrated over $dx$) consists of 3 contributions: The kinetic energy from continuum states, subtracting the asymptotic value
\begin{equation}
E_{\rm loc}^{(1)} = N \int_{-\infty}^{\infty} dx \int_{1/\Lambda}^{\Lambda}\frac{d\zeta}{2\pi}   \frac{\zeta^2+1}{2 \zeta^2}
 \left( - i{\psi}_{\zeta}^{\dagger} \gamma_5 \partial_x 
\psi_{\zeta} +  \frac{(1-\zeta^2)^2}{2\zeta(1+\zeta^2)} \right),
\label{2.29}
\end{equation}
the subtracted potential energy
\begin{equation}
E_{\rm loc}^{(2)} = - \frac{N}{2Ng^2} \int_{- \infty}^{\infty} dx \left( S^2-1 \right),
\label{2.30}
\end{equation}
and the kinetic energy from the discrete states
\begin{equation}
E_{\rm loc}^{(3)} = - i N \int_{-\infty}^{\infty} dx    \left( {\psi}^{(1)\dagger} \gamma_5 \partial_x 
\psi^{(1)} + \nu {\psi}^{(2)\dagger} \gamma_5 \partial_x  \psi^{(2)}    \right).
\label{2.31}
\end{equation}
Using the vacuum gap equation
\begin{equation}
\frac{\pi}{Ng^2} = \ln \Lambda
\label{2.32}
\end{equation}
and performing all integrations analytically, one finds as usual that $E_{\rm loc}^{(2)}$ cancels exactly the logarithmic divergence in 
$E_{\rm loc}^{(1)}$. The finite part of $E_{\rm loc}^{(1)}$ combines with $E_{\rm loc}^{(3)}$ to a complicated expression multiplied by a factor
\begin{equation}
\nu-1 + \frac{2Q}{1+Q^2} + \frac{4iQ\ln Z}{\pi(1+Q^2)},
\label{2.33}
\end{equation}
which vanishes owing to the self-consistency condition. Hence the local energy part of the energy vanishes. 
The total mass of the breather is then completely determined by the fermion phase shifts \cite{L2,L5}, which are identical to those of a baryon
with the same value of $Z$ or $\epsilon$. We conclude that the mass of the breather with parameters ($b,\epsilon$) is the same
as the baryon mass with parameter $\epsilon$,
\begin{equation}
M_{\rm breather}(\epsilon,b) = \frac{2\epsilon N}{\pi \sqrt{1+\epsilon^2}},
\label{2.34}
\end{equation}
in agreement with \cite{L2}.
Thus the breathers shown in Figs.~\ref{Fig2} and \ref{Fig3} have a common mass ($M/N=0.56941003$), as do the breathers
shown in Figs.~\ref{Fig4} and \ref{Fig5} ($M/N=0.63661912$).
Independence of the mass on the breather amplitude is counter-intuitive at first sight. However, we have to remember that the 
baryon number depends on $Q$ and decreases with growing amplitude, so that there are competing effects.  
If we compare the mass of the breather with the mass of a baryon with the same fermion number, using Eq.~(\ref{2.10}), we indeed find
that the breather gets heavier with increasing amplitude,
\begin{equation}
M_{\rm baryon} = \frac{2N}{\pi}\sin \left(\frac{\pi \nu}{2}\right), \quad M_{\rm breather} = \frac{2N}{\pi} \sin \left(\frac{\pi(\lambda-1+\nu)}
{2\lambda} \right).
\label{2.35}
\end{equation}
At fixed $\nu$, the ratio $M_{\rm breather}/M_{\rm baryon}$ increases monotonically with $\lambda$. 
Clearly, the breather mass always stays below the kink-antikink threshold $2N/\pi$, a precondition for its stability.

\subsection{Exponential ansatz method for single breather}\label{sect2d}

Scattering problems involving baryons and multi-baryon states have been solved recently by means of a joint ansatz for the scalar TDHF potential
and the Dirac spinors \cite{L8,L9}. Assuming that the self-consistent potentials are transparent and working with rational functions
of certain basis  exponentials, it was possible 
to solve the Dirac equation purely algebraically. This method has the potential to handle the breather-breather scattering problem as well.
In the present section, we cast the results for the single breather into a form well suited for such a scattering calculation.
To this end, we have to express the single breather potential and spinors in terms of basis exponentials. To account for the new fact 
that both hyperbolic and trigonometric functions appear in $S$, Eq.~(\ref{2.2}), we evidently must complexify the basis exponentials. 

This section is also a preparation for tackling the breather-breather scattering problem in Sect.~\ref{sect3}. There, we obviously will have
to deal with 2 different breathers, moving at different velocities. To keep the notation consistent, it is therefore advisable to introduce
breather labels $i=1,2$ right away, even if it renders the notation in the present subsection more cumbersome. To distinguish the breather 
labels from many other subscripts or superscripts, we will use superscripts $(i)$ with parentheses. 

So far, we have only considered the breather at rest. Boosting the breather to different 
velocities is straightforward in light cone coordinates. We will indicate the necessary modifications introduced by the boost as we go along.    

We introduce as basis exponentials for breather $i$ in the notation of subsection~\ref{sect2b}
\begin{equation}
V_1^{(i)} = \exp \left( \frac{i \eta_iz}{2 Z_i}- \frac{i Z_i\bar{z}}{2 \eta_i}\right), \quad V_2^{(i)} = V_1^{(i)*}.
\label{2.36}
\end{equation}
The boost parameter $\eta_i$ depends on the breather velocity $v_i$, see Eq.~(\ref{2.18}).
The breather potential (\ref{2.2}) can be written as the following rational function of $V_1^{(i)}, V_2^{(i)}$,
\begin{eqnarray}
S^{(i)} & = & \frac{{\cal N}^{(i)}}{{\cal D}^{(i)}},
\nonumber \\
{\cal N}^{(i)} & = & 1 + a_{11}^{(i)} (V_1^{(i)})^2 + a_{12}^{(i)} V_1^{(i)} V_2^{(i)} + a_{22}^{(i)} (V_2^{(i)})^2 + a_{1122}^{(i)} (V_1^{(i)})^2 (V_2^{(i)})^2,
\nonumber \\
{\cal D}^{(i)} & = & 1 + b_{11}^{(i)} (V_1^{(i)})^2 + b_{12}^{(i)} V_1^{(i)} V_2^{(i)} + b_{22}^{(i)} (V_2^{(i)})^2 + b_{1122}^{(i)} (V_1^{(i)})^2 (V_2^{(i)})^2,
\label{2.37}
\end{eqnarray}
with
\begin{eqnarray}
a_{11}^{(i)} & = & a_{22}^{(i)}  =  - b_{11}^{(i)}  = -b_{22}^{(i)} = \frac{Z_i^2-1}{Z_i^2+1}\frac{Q_i^2-1}{Q_i^2+1} ,
\nonumber \\
a_{12}^{(i)} & = & \frac{Z_i^4+1}{Z_i^2}\frac{2Z_i}{Z_i^2+1}\frac{2Q_i}{Q_i^2+1}  , \quad b_{12}^{(i)} = 2 \frac{2Z_i}{Z_i^2+1}\frac{2Q_i}{Q_i^2+1},
\nonumber \\
a_{1122}^{(i)} & = & b_{1122}^{(i)} = 1.
\label{2.38}
\end{eqnarray}
The ansatz for the continuum spinors (assuming a transparent potential) follows exactly the strategy
used earlier for baryons \cite{L8}, 
\begin{equation}
\psi_{\zeta}^{(i)} = \frac{\exp \left( \frac{i \zeta \bar{z}}{2}- \frac{iz}{2\zeta}\right)}{\sqrt{1+\zeta^2} {\cal D}^{(i)}} \left( \begin{array}{c}
\zeta {\cal N}_1^{(i)} \\ - {\cal N}_2^{(i)} \end{array} \right),
\label{2.39}
\end{equation}
with
\begin{eqnarray}
{\cal N}_1^{(i)} & = & 1 + c_{11}^{(i)} (V_1^{(i)})^2 + c_{12}^{(i)} V_1^{(i)} V_2^{(i)} + c_{22}^{(i)} (V_2^{(i)})^2 + c_{1122}^{(i)} (V_1^{(i)})^2 (V_2^{(i)})^2,
\nonumber \\
{\cal N}_2^{(i)} & = & 1 + d_{11}^{(i)} (V_1^{(i)})^2 + d_{12}^{(i)} V_1^{(i)} V_2^{(i)} + d_{22}^{(i)} (V_2^{(i)})^2 + d_{1122}^{(i)} (V_1^{(i)})^2 (V_2^{(i)})^2.
\label{2.40}
\end{eqnarray}
The free spinor has been pulled out so that all polynomials start with a 1. 
When inserted into the Dirac equation, this ansatz leads to an overdetermined algebraic problem. 
We find  a unique solution with coefficients
\begin{eqnarray}
c_{11}^{(i)} & = & - d_{11}^{(i)} = - \frac{\zeta \eta_i+Z_i}{\zeta \eta_i-Z_i} a_{11}^{(i)},
\nonumber \\
c_{22}^{(i)} & = & -d_{22}^{(i)} = - \frac{\zeta \eta_iZ_i -1}{\zeta \eta_i Z_i+1} a_{11}^{(i)},
\nonumber \\
c_{1122}^{(i)} & = & d_{1122}^{(i)} = \frac{\zeta \eta_i+Z_i}{\zeta \eta_i-Z_i} \frac{\zeta \eta_iZ_i -1}{\zeta \eta_i Z_i+1},
\nonumber \\
c_{12}^{(i)} & = & -  \frac{Z_i^4+1-2Z_i^2 \zeta^2 \eta_i ^2}{2Z_i(\zeta \eta_i - Z_i)( \zeta \eta_i Z_i+1)} b_{12}^{(i)},
\nonumber \\
d_{12}^{(i)} & = & - \frac{2 Z_i^2- \zeta^2 \eta_i^2(Z_i^4+1)}{2Z_i(\zeta \eta_i - Z_i)( \zeta \eta_i Z_i+1)} b_{12}^{(i)}.
\label{2.41}
\end{eqnarray}
Since the potential is transparent, the fermion-breather transmission amplitude $T^{(i)}$ is a pure phase factor which can be read off from 
$c_{1122}^{(i)}$ or $d_{1122}^{(i)}$.
Interestingly, the result is independent of $Q_i$,
\begin{equation}
T^{(i)}= \frac{\zeta \eta_i+Z_i}{\zeta \eta_i-Z_i} \frac{\zeta \eta_iZ_i -1}{\zeta \eta_i Z_i+1},
\label{2.42}
\end{equation} 
and hence coincides with the result for a baryon with the same values of $Z_i, \eta_i$. It is worthwhile to interpret the pole structure
of $T^{(i)}$ in physical terms. The singularities of $T^{(i)}$ in the complex $\zeta$-plane are located at 
\begin{equation}
\zeta= \frac{Z_i}{\eta_i}, \qquad \zeta = - \frac{1}{\eta_i Z_i}.
\label{2.42a}
\end{equation}
Going back to ordinary coordinates via Eq.~(\ref{2.17}), this yields 
\begin{equation}
k  = \frac{Z_i^2-\eta_i^2}{2 \eta_i Z_i},  \quad  \omega = - \frac{Z_i^2+\eta_i^2}{2 \eta_i Z_i}
\label{2.42b}
\end{equation}
for $\zeta= Z_i/\eta_i$ and
\begin{equation}
k   =   \frac{\eta_i^2 Z_i^2 -1}{2\eta_i Z_i},  \quad  \omega=  \frac{Z_i^2+\eta_i^2}{2 \eta_i Z_i}
\label{2.42c}
\end{equation}
for $\zeta= -1/\eta_iZ_i$, respectively. In the rest frame, $\eta_i = 1$ and $\omega$ agrees with the quasi-energy $\pm 1/\sqrt{1+\epsilon^2}$ of the breather bound states (or the energy eigenvalue in the baryon case).
If we associate the imaginary momentum $k= i \epsilon/\sqrt{1+\epsilon^2}$ with the bound states in the rest frame, it is easy to check that Eqs.~(\ref{2.42b},\ref{2.42c}) are just the result
of boosting the 2-vectors $(k,\pm \omega)$ to a moving frame.

We now turn to the bound state spinors for the two discrete states which we shall label with double superscripts, $\psi^{(i,j)}$ 
(the $j$-th bound state of breather $i$). In the case of the DHN baryon, the easiest way to get the 
bound state spinors is to compute the residue of the continuum spinor at the poles of the transmission amplitude in the complex $\zeta$-plane. 
This works here as well (the poles are at the same location), but the resulting states are not orthogonal and therefore 
not directly suited for the TDHF calculation. The difference comes about because in the breather case, the discrete states are not eigenstates 
of the Hamiltonian, so that there is an ambiguity which linear combination one should choose, as also noticed by DHN \cite{L2}. We impose
two criteria: We demand that the bound state spinors $\psi^{(i,j)}$ satisfy the orthogonality and normalization conditions
\begin{equation}
\int dx \psi^{(i,j)\dagger}(x,t) \psi^{(i,k)}(x,t) = \delta_{jk},
\label{2.43}
\end{equation}
and that the (local) scalar condensate matrix is diagonal, 
\begin{equation}
\bar{\psi}^{(i,j)}(x,t) \psi^{(i,k)}(x,t) = \delta_{jk} \bar{\psi}^{(i,j)}(x,t)\psi^{(i,j)}(x,t).
\label{2.44}
\end{equation}
If one insists on Eqs.~(\ref{2.43}, \ref{2.44}), one finds that the spinors are unique up to overall phases and
interchanging the state labels (1,2). This choice then enables us to use the TDHF equation, Eq.~(\ref{2.1}), in the standard form
for the discrete states as well. Our results agree with DHN who have reasoned somewhat differently in the semi-classical path integral 
approach. The resulting bound state spinors can be represented in the form
\begin{eqnarray}
\psi^{(i,j)} & = & \frac{C_0^{(i)}}{{\cal D}^{(i)}}  \left( \begin{array}{c} {\cal N}_1^{(i,j)} \\ \eta_i {\cal N}_2^{(i,j)} \end{array} \right)  \quad (j=1,2),
\nonumber \\
{\cal N}_1^{(i,j)} & = & e_1^{(i,j)} V_1^{(i)} + e_2^{(i,j)} V_2^{(i)} +e_{112}^{(i,j)} (V_1^{(i)})^2 V_2^{(i)} + e_{122}^{(i,j)} V_1^{(i)} (V_2^{(i)})^2,
\nonumber \\
{\cal N}_2^{(i,j)} & = & f_1^{(i,j)} V_1^{(i)} + f_2^{(i,j)} V_2^{(i)} +f_{112}^{(i,j)} (V_1^{(i)})^2 V_2^{(i)} + f_{122}^{(i,j)} V_1^{(i)} (V_2^{(i)})^2.
\label{2.45}
\end{eqnarray}
The factor $\eta_i$ in the lower spinor component accounts for the different transformation properties of left-handed and
right-handed spinors under Lorentz boosts, leading to more symmetric coefficients.  
These coefficients are found to be
\begin{eqnarray}
e_1^{(i,1)} & = & f_2^{(i,2)} = e_{122}^{(i,2)} = - f_{112}^{(i,1)} =  Z_i(1-Q_i),
\nonumber \\
e_2^{(i,1)} & = & f_1^{(i,2)} = e_{112}^{(i,2)} = -f_{122}^{(i,1)} = 1+Q_i,
\nonumber \\
e_{112}^{(i,1)} & = & - f_1^{(i,1)} = e_2^{(i,2)} = f_{122}^{(i,2)} = -(1-Q_i),
\nonumber \\
e_{122}^{(i,1)} & = & -f_2^{(i,1)} = e_1^{(i,2)} = f_{112}^{(i,2)} = Z_i(1+Q_i).
\label{2.46}
\end{eqnarray}
The normalization factor can be chosen as real and positive,
\begin{equation}
C_0^{(i)} = \left(-  \frac{i Q_i(Z_i^2-1)}{2\eta_i (Z_i^2+1)(Q_i^2+1)}\right)^{1/2}.
\label{2.47}
\end{equation}
All the results of this section reduce to the corresponding single baryon results if we set $Q_i=-1$. The (real) product $V_1^{(i)}V_2^{(i)}$ 
can be identified with $U^{(i)}$ of the static baryon in that case. The spinors agree up to irrelevant overall phase factors.  
$\psi^{(i,1)}$ goes over into the negative, $\psi^{(i,2)}$ into the positive energy bound state spinor. Both in the baryon 
and the breather case, $\psi^{(i,1)}$ and $\psi^{(i,2)}$ are related by charge conjugation which reads
\begin{equation}
\psi^{(i,2)} = \gamma_5 \psi^{(i,1)*}
\label{2.48}
\end{equation}
in our Dirac basis. 

Consider the computation of the fermion density next. Assuming the lower bound state to be fully occupied and an occupation fraction $\nu$ for the 
upper bound state, the total fermion density of breather $i$ is given by 
\begin{equation}
\rho = \nu N \psi^{(i,1)\dagger} \psi^{(i,1)}.
\label{2.48a}
\end{equation}
The fermion density of the lower bound state is cancelled exactly against the fermion density induced in the Dirac sea. If one simply inserts the bound state spinors (\ref{2.45})
into expression (\ref{2.48a}), one gets a rather intransparent result. A more convenient way of accessing the fermion density is as follows. Since the vector
current $j^{\mu}$ for each single particle state is conserved, it can be represented in terms of a pseudoscalar field $P$ as
\begin{equation}
\partial_{\mu} j^{\mu} = 0 \quad \longrightarrow \quad j^{\mu} = \epsilon^{\mu \nu} \partial_{\nu} P.
\label{2.48b}
\end{equation}
We use the convention
\begin{equation}
\left( \epsilon^{\mu \nu} \right) = \left( \begin{array}{rr} 0 & -1 \\ 1 & \ 0 \end{array} \right), \quad 
\left( \epsilon_{\mu \nu} \right) = \left( \begin{array}{rr} 0 & 1 \\ -1 & \ 0 \end{array} \right).
\label{2.48c}
\end{equation}
We note in passing that the axial current can then be expressed as follows,
\begin{equation}
j_5^{\mu} = \epsilon^{\mu \nu} j_{\nu} = \partial^{\mu} P,
\label{2.48d}
\end{equation}
however it is not conserved in the GN model.
Clearly, $P$ is only defined up to an additive constant. By a judicious choice of this constant, one gets a very simple expression for $P$ of breather $i$,
\begin{equation}
P^{(i)} = \frac{(V_1^{(i)} V_2^{(i)})^2-1}{2 {\cal D}^{(i)}}.
\label{2.48e}
\end{equation}
The best way of computing the fermion density and the fermion current is to insert this expression into (\ref{2.48b}), i.e.,
\begin{equation}
\rho^{(i)} = \partial_x P^{(i)}, \quad j^{(i)} = - \partial_t P^{(i)}. 
\label{2.48e1}
\end{equation}
The normalization condition then reduces to
\begin{equation}
\int_{-\infty}^{\infty} dx \rho^{(i)}= P^{(i)}(x=\infty) - P^{(i)}(x=-\infty) = 1.
\label{2.48f}
\end{equation}
Indeed $P^{(i)}$ of Eq.~(\ref{2.48e}) has the shape of a kink evolving from $-1/2$ to 1/2 with our choice of the additive constant.

Finally, let us recall how the self-consistency condition (\ref{2.5}) arises, dropping the breather label for the moment and returning to
the original parameters ($\epsilon,b$). The scalar density from the continuum spinors yields 
\begin{eqnarray}
\bar{\psi}_{\zeta} \psi_{\zeta} & = &  (\bar{\psi}_{\zeta} \psi_{\zeta})_1 +  (\bar{\psi}_{\zeta} \psi_{\zeta})_2
\nonumber \\
(\bar{\psi}_{\zeta} \psi_{\zeta})_1 & = & - \frac{2\zeta}{\zeta^2+1} S
\label{2.49}
\end{eqnarray}
Upon using the vacuum gap equation, the first term gives self-consistency. The 2nd term can be summed over the Dirac sea 
(integration over $d\zeta$ with the
appropriate measure) with the result
\begin{equation}
\int_0^{\infty} \frac{d\zeta}{2\pi} \left( \frac{\zeta^2+1}{2\zeta^2} \right) (\bar{\psi}_{\zeta} \psi_{\zeta} )_2 = - \frac{\epsilon b}
{(1+\epsilon^2)}\left(1 - \frac{2}{\pi} \arctan \epsilon \right) \frac{2V_1 V_2}{\cal D}
\label{2.50}
\end{equation}
The discrete spinors contribute
\begin{equation}
(\bar{\psi} \psi)^{(1,2)} = \mp \frac{\epsilon}{\sqrt{1+\epsilon^2}} \frac{2 V_1 V_2}{\cal D}
\label{2.51}
\end{equation}
to the condensate.
Denoting the fermion numbers of states $1,2$ by $N, \nu N$, respectively, the condition that (\ref{2.46}) and (\ref{2.47}) cancel 
reproduces exactly the DHN result, Eq.~(\ref{2.5}).

The breather is a type II TDHF solution, like the DHN baryon, in the classification scheme of Ref.~\cite{L6}.
Incidentally, one can easily check that the breather is in general not a self-consistent solution for the massive GN model. 
In the massive model only the baryon limit $Q=-1$ is self-consistent \cite{L16}. Hence we do not expect the DHN breather to survive in the
nonrelativistic limit. It is a genuine relativistic object, like the kink or antikink.

\section{Breather-breather scattering}\label{sect3}

Here we follow almost literally the strategy proven successful in previous studies of baryon-baryon scattering \cite{L8,L9}. We write
down an ansatz for the two-breather problem, starting from the known single breather solution.  Next, ``reducible" parameters are
determined from the asymptotics of the incoming and outgoing breathers. The remaining ``irreducible" parameters can be found 
by solving the Dirac equation algebraically. This procedure yields a reflectionless, time dependent scalar potential, together with the
continuum and bound state spinors. We then check the self-consistency of the solution and compute the fermion density. The final results
will be illustrated with a few examples. Technically, the present problem is more involved than baryon-baryon scattering which it
generalizes to scattering of collectively excited hadrons. We believe that the problem is nevertheless worth the effort. 
When going from baryon-baryon scattering to scattering of any number of composite bound states, we observed a kind of factorization
which allowed us to reduce the dynamical multi-baryon problem to baryon-baryon scattering \cite{L9}. Similarly, we expect the solution
of the breather-breather scattering problem to be the key element in a future study of scattering of any number of breathers,
baryons and composites thereof.

\subsection{Ansatz and asymptotic conditions}\label{sect3a}

We are now dealing with two (boosted) breathers with unavoidable notational complications. In Sect.~\ref{sect2}, the basis 
exponentials referring to breather $i$ were denoted by $V_{1,2}^{(i)}$, see Eq.~(\ref{2.36}). The same exponentials will be
used in the scattering problem for breathers 1 and 2. However it turns out that both the notation and the bookkeeping are
somewhat simpler if we label the exponentials of breather 1 by $V_{1,2}$ and those of breather 2 by $V_{3,4}$ from now on,
\begin{eqnarray}
V_1 & = &  \exp \left(  \frac{i\eta_1 z}{2Z_1}- \frac{i Z_1\bar{z}}{2 \eta_1}\right), \quad V_2 = V_1^*,
\nonumber \\
V_3 & = &  \exp \left(  \frac{i\eta_2 z}{2Z_2} - \frac{iZ_2\bar{z}}{2 \eta_2}\right), \quad V_4 = V_3^*.
\label{3.1}
\end{eqnarray}
We shall always assume $v_1>v_2$ in the following. The scalar potential is written as 
\begin{equation}
S  =  \frac{\cal N}{\cal D},
\label{3.2}
\end{equation}
where numerator and denominator are multivariate polynomials of all four exponentials $V_k$. The degree 
of these polynomials is simply determined by 
multiplying the scalar potentials of two distinct breathers and keeping all the terms appearing there, but with
unknown coefficients. Thus the basic structure of ${\cal N}$ and ${\cal D}$ consists of 25 terms generated from
\begin{equation}
(1+V_1^2 + V_1 V_2 + V_2^2  + V_1^2 V_2^2)(1+V_3^2+V_3V_4 + V_4^2 + V_3^2V_4^2).
\label{3.3}
\end{equation}
The ansatz for the numerator ${\cal N}$ then reads 
\begin{eqnarray}
{\cal N} & = & 1 + a_{11} V_1^2 + a_{22} V_2^2 +a_{33} V_3^2 + a_{44} V_4^2 +a_{12} V_1 V_2 +a_{34} V_3 V_4 
\nonumber \\
& & + a_{1122} V_1^2 V_2^2+ a_{1133} V_1^2 V_3^2 + a_{1144} V_1^2 V_4^2 + a_{2233} V_2^2 V_3^2 + a_{2244} V_2^2 V_4^2 
+ a_{3344} V_3^2 V_4^2
\nonumber \\
& & + a_{1233} V_1 V_2 V_3^2 + a_{1244} V_1 V_2 V_4^2 + a_{1134} V_1^2 V_3 V_4 + a_{2234} V_2^2 V_3 V_4 + a_{1234} V_1 V_2 V_3 V_4
\nonumber \\
& & + a_{112233} V_1^2V_2^2V_3^2+ a_{112244} V_1^2V_2^2 V_4^2 + a_{113344} V_1^2 V_3^2 V_4^2 + a_{223344} V_2^2 V_3^2 V_4^2
\nonumber \\
&  &  + a_{112234} V_1^2 V_2^2 V_3 V_4 + a_{123344} V_1 V_2 V_3^2 V_4^2
+ a_{11223344} V_1^2V_2^2V_3^2V_4^2.
\label{3.4}
\end{eqnarray}
The denominator ${\cal D}$ has the same structure as ${\cal N}$ with all $a$-coefficients replaced by $b$-coefficients.
Similarly, the continuum spinors are parametrized as
\begin{equation}
\psi_{\zeta} = \frac{\exp \left( \frac{i \zeta \bar{z}}{2}- \frac{iz}{2\zeta}\right)}{\sqrt{1+\zeta^2} {\cal D}} \left( \begin{array}{c}
\zeta {\cal N}_1 \\ - {\cal N}_2 \end{array} \right).
\label{3.5}
\end{equation}
Here, ${\cal N}_1,{\cal N}_2$ have the same polynomial form as ${\cal N}$, Eq.~(\ref{3.4}), with all $a$-coefficients replaced by 
$c$- and $d$-coefficients, respectively. Like in baryon-baryon scattering we expect four distinct bound states, two for each breather.
They will be labelled by 1 and 2 (for breather 1) and 3 and 4 (for breather 2), where the identification of the bound state
with a particular breather always refers to the asymptotic region.
The polynomial structure for bound states asymptotically belonging to breather 1 can be obtained by multiplying the numerator of 
the bound state of breather 1 with the numerator (or denominator) of $S$ of breather 2 (20 terms), 
\begin{equation}
(V_1+V_2)(1+V_1V_2)(1+V_3^2 + V_3V_4 + V_4^2 + V_3^2 V_4^2).
\label{3.6}
\end{equation} 
This suggests the following ansatz for the first bound state spinor,
\begin{equation}
\psi^{(1)} =\frac{C^{(1)}}{\cal D} \left( \begin{array}{r} {\cal N}_1^{(1)} \\ \eta_1 {\cal N}_2^{(1)} \end{array} \right).
\label{3.7}
\end{equation}
Here,
\begin{eqnarray}
{\cal N}_{1}^{(1)} & = & e_1^{(1)} V_1 + e_2^{(1)} V_2 + e_{112}^{(1)} V_1^2 V_2  + e_{122}^{(1)} V_1 V_2^2 + e_{133}^{(1)} V_1 V_3^2 + 
e_{134}^{(1)} V_1 V_3 V_4  
\nonumber \\
& & + e_{144}^{(1)} V_1 V_4^2 + e_{233}^{(1)} V_2 V_3^2 + e_{234}^{(1)} V_2 V_3 V_4 + e_{244}^{(1)} V_2 V_4^2
\nonumber \\
& & + e_{11233}^{(1)} V_1^2 V_2 V_3^2 +  e_{11234}^{(1)} V_1^2 V_2 V_3 V_4 + e_{11244}^{(1)} V_1^2 V_2 V_4^2 + e_{12233}^{(1)} V_1 V_2^2 V_3^2  
\nonumber \\
& & + e_{12234}^{(1)} V_1 V_2^2 V_3 V_4 +e_{12244}^{(1)} V_1 V_2^2 V_4^2 + e_{13344}^{(1)} V_1  V_3^2 V_4^2 + e_{23344}^{(1)} V_2 V_3^2  V_4^2
\nonumber \\
& & + e_{1123344}^{(1)} V_1^2 V_2 V_3^2 V_4^2 + e_{1223344}^{(1)} V_1 V_2^2 V_3^2 V_4^2,
\label{3.8}
\end{eqnarray}
and ${\cal N}_2^{(1)}$ has the same structure with all $e^{(1)}$-coefficients replaced by $f^{(1)}$-coefficients.
Similarly, bound states related to breather 2 have the structure generated by a bound state of breather 2 and the
numerator (or denominator) of $S$ of breather 1,
\begin{equation}
(V_3+V_4)(1+V_3V_4)(1+V_1^2+V_1V_2 + V_2^2 + V_1^2V_2^2).
\label{3.9}
\end{equation}
The 3rd bound state spinor then becomes 
\begin{equation}
\psi^{(3)} =\frac{C^{(3)}}{\cal D} \left( \begin{array}{r} {\cal N}_1^{(3)} \\ \eta_2 {\cal N}_2^{(3)} \end{array} \right)
\label{3.10}
\end{equation}
with
\begin{eqnarray}
{\cal N}_1^{(3)} & = & e_3^{(3)} V_3+ e_4^{(3)}V_4 + e_{113}^{(3)}V_1^2V_3 + e_{114}^{(3)}V_1^2V_4 + e_{123}^{(3)}V_1V_2V_3+ e_{124}^{(3)}
V_1V_2V_4
\nonumber \\
& & + e_{223}^{(3)}V_2^2V_3 + e_{224}^{(3)}V_2^2V_4+ e_{334}^{(3)}V_3^2V_4 + e_{344}^{(3)}V_3V_4^2
\nonumber \\
& & + e_{11223}^{(3)}V_1^2V_2^2V_3 + e_{11224}^{(3)}V_1^2V_2^2V_4 + e_{11334}^{(3)}V_1^2V_3^2V_4 + e_{11344}^{(3)}V_1^2V_3V_4^2
\nonumber \\
& & + e_{12334}^{(3)}V_1V_2V_3^2V_4 + e_{12344}^{(3)}V_1V_2V_3V_4^2 + e_{22334}^{(3)}V_2^2V_3^2V_4 + e_{22344}^{(3)}V_2^2V_3V_4^2
\nonumber \\
& & + e_{1122334}^{(3)}V_1^2V_2^2V_3^2V_4 + e_{1122344}^{(3)}V_1^2V_2^2V_3V_4^2.
\label{3.11}
\end{eqnarray}
Once again ${\cal N}_2^{(3)}$ has the same structure with all $e^{(3)}$-coefficients replaced by $f^{(3)}$-coefficients.
The two remaining bound state spinors can simply be obtained by charge conjugation,
\begin{equation}
\psi^{(2)} =\gamma_5 \psi^{(1)*}, \quad \psi^{(4)} = \gamma_5 \psi^{(3)*}.
\label{3.12}
\end{equation}
Our labelling is such that in the baryon limit, bound states $1,3$ become the negative, $2,4$ the positive energy discrete states.
This ansatz leaves us altogether with a large number of coefficients to be determined,
namely 2$\times$25 for $S$, 2$\times$25 for the continuum spinors and another 2$\times$20 for each bound state spinor. 
Even if we make use of Eq.~(\ref{3.12}) we need to determine as many as 180 coefficients.

The next step consists in reducing this large number by exploiting the asymptotic conditions in the initial and final 
states where the breathers are well separated. This is again analogous to what has been done for baryons in Refs.~\cite{L8,L9}.
Consider $S$ first. Initial and final breathers can be projected out by letting $(V_1,V_2)$ or $(V_3,V_4)$
go simultaneously to 0 or infinity. The relationship between $S(V_1,V_2,V_3,V_4)$ for breather-breather scattering and $S(V_1,V_2),
S(V_3,V_4)$ for the two individual breathers is
\begin{eqnarray}
\lim_{V_3,V_4\to 0}S(V_1,V_2,V_3,V_4) & = & S(V_1,V_2),
\nonumber \\
\lim_{V_3,V_4\to \infty}S(V_1,V_2,V_3,V_4) & = & S(\sigma_1V_1,\sigma_2V_2),
\nonumber \\
\lim_{V_1,V_2\to 0}S(V_1,V_2,V_3,V_4) & = & S(\sigma_3 V_3, \sigma_4 V_4),
\nonumber \\
\lim_{V_1,V_2\to \infty}S(V_1,V_2,V_3,V_4) & = & S(V_3,V_4).
\label{3.13}
\end{eqnarray}
\begin{figure}[h]
\begin{center}
\epsfig{file=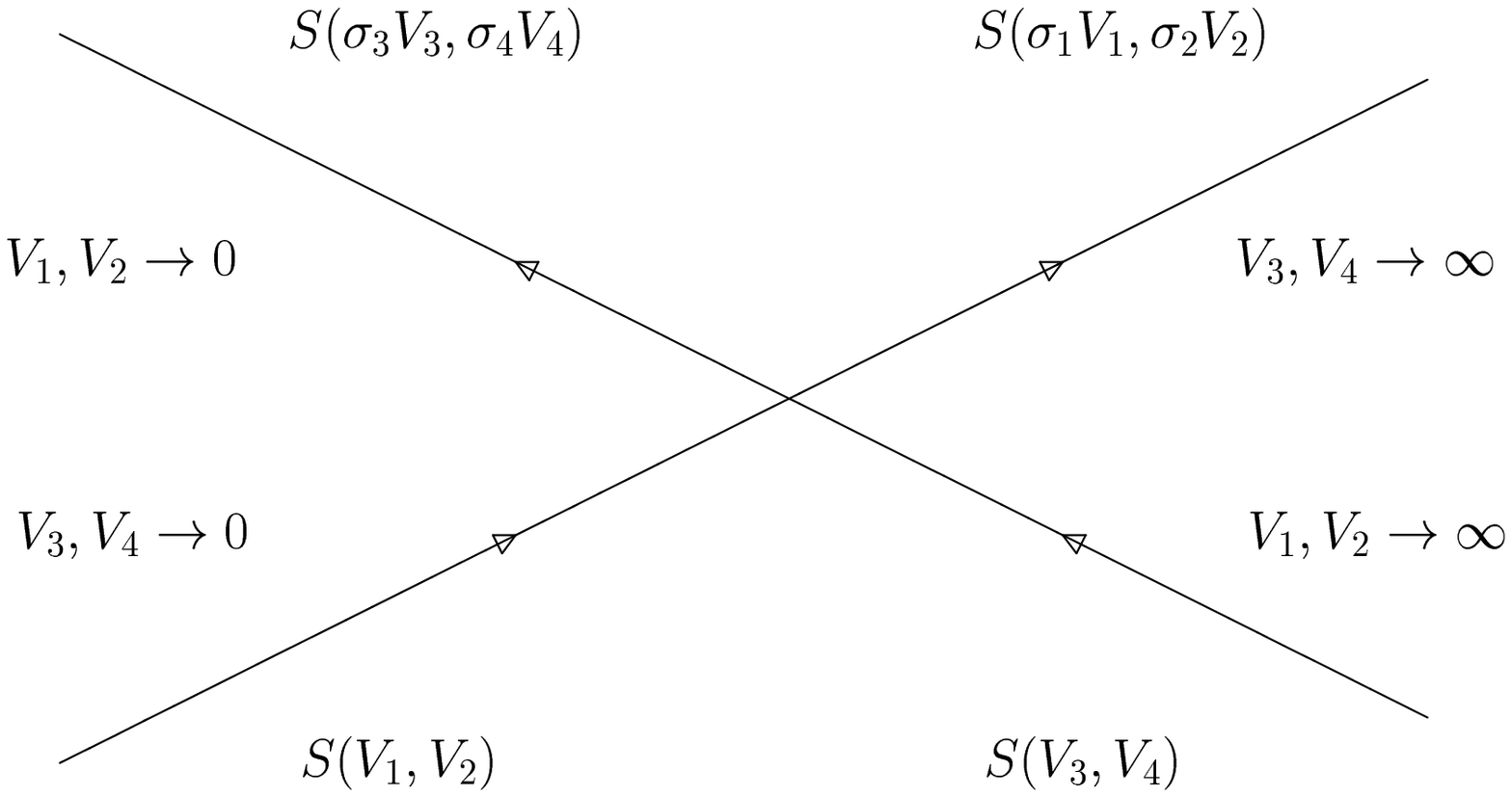,angle=270,width=8cm}
\caption{Asymptotic reduction of scalar potential for breather-breather scattering. Time runs in the vertical, space in the horizontal direction.
Outgoing breathers 1 and 2 experience a time delay and a shift in their phase, as indicated by the complex factors $\sigma_i$.}
\label{Fig6}
\end{center}
\end{figure}
As illustrated in Fig.~\ref{Fig6}, the (complex) factors $\sigma_i$ account for the fact that during the collision the breather undergoes a 
time delay and a possible change in phase. Thus they encode the whole asymptotic scattering information.
Since $S$ is real, they have to satisfy
\begin{equation}
\sigma_2 = \sigma_1^*, \quad \sigma_4 = \sigma_3^*.
\label{3.14}
\end{equation}
In the baryon limit, $U_1=V_1V_2, U_2=V_3V_4$ and the $\sigma_i$ are related to the (real) time delay factors $\delta_{12}$  
introduced in \cite{L8},
\begin{equation}
\frac{1}{\delta_{12}} = \sigma_1 \sigma_2 = |\sigma_1|^2, \quad \frac{1}{\delta_{21}} = \sigma_3 \sigma_4 = |\sigma_3|^2.
\label{3.15}
\end{equation}
The condition $\delta_{12} \delta_{21}=1$ found for baryons has been confirmed here for breathers in the form
\begin{equation}
\sigma_1 \sigma_2 \sigma_3 \sigma_4 = 1.
\label{3.16}
\end{equation}
\begin{figure}[h]
\begin{center}
\epsfig{file=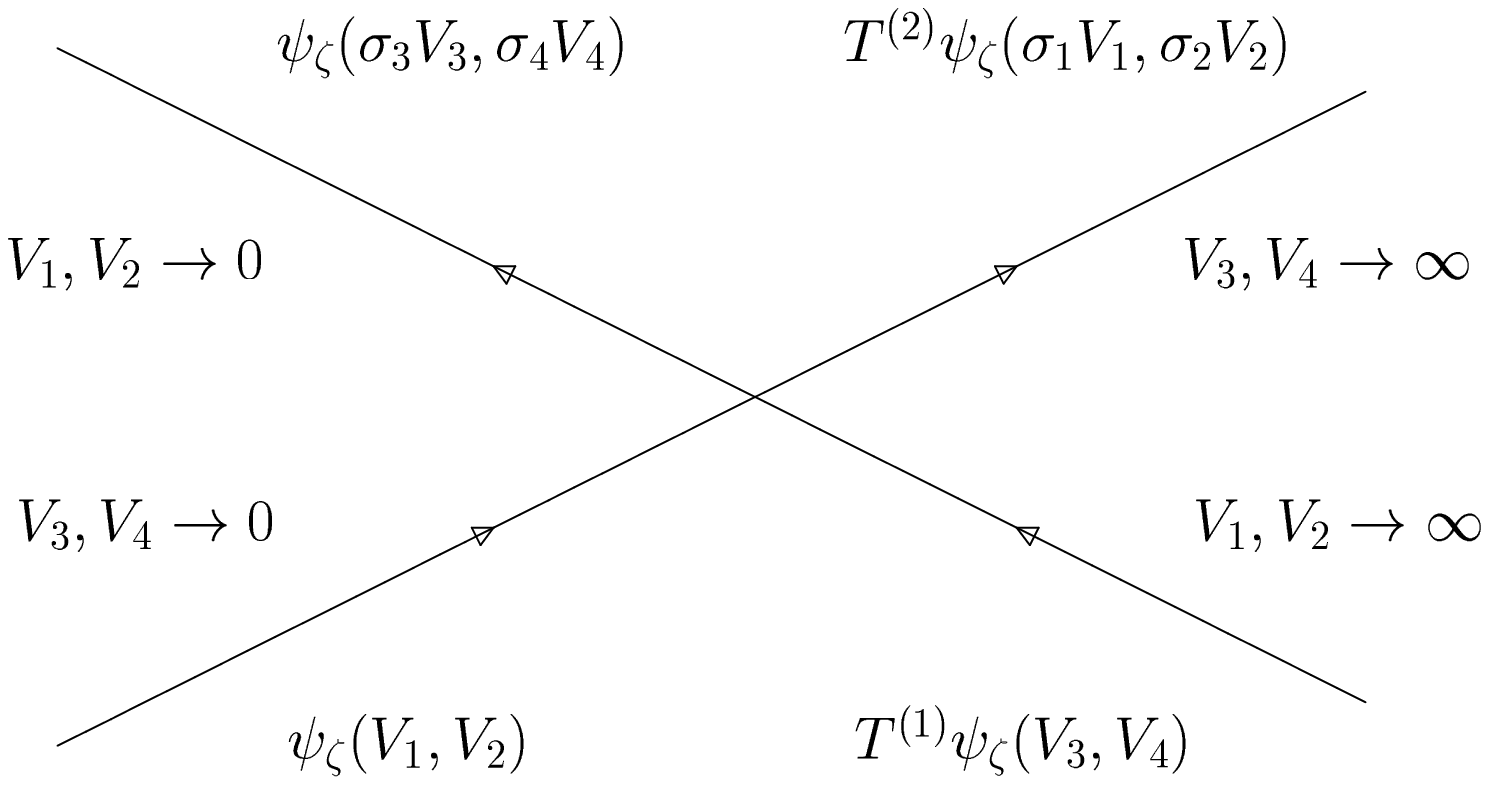,angle=270,width=8cm}
\caption{Like Fig.~\ref{Fig6}, but for continuum spinors. The incoming spinor 2 and the outgoing spinor 1 acquire transmission 
amplitudes $T^{(1,2)}$ from scattering on the other breather, since they are incident from the left in this figure.}
\label{Fig7}
\end{center}
\end{figure}

The asymptotic reduction for the continuum spinors $\psi_{\zeta}$ is similar to the one for $S$, except that one has to take into account 
that the spinor acquires a transmission amplitude in some cases, see Fig.~\ref{Fig7}. Since the scattering wave functions are defined to be 
incident from the left, the transmission amplitudes affect the two breathers on the right side of Fig.~\ref{Fig7}, namely incoming breather 2
and outgoing breather 1. Denoting the transmission amplitude of the spinor due to breather $i$ by $T^{(i)}$ as in Sect.~\ref{sect2d}, we get
\begin{eqnarray}
\lim_{V_3,V_4\to 0}\psi_{\zeta}(V_1,V_2,V_3,V_4) & = & \psi_{\zeta}(V_1,V_2),
\nonumber \\
\lim_{V_3,V_4\to \infty}\psi_{\zeta}(V_1,V_2,V_3,V_4) & = & T^{(2)} \psi_{\zeta}(\sigma_1 V_1,\sigma_2 V_2),
\nonumber \\
\lim_{V_1,V_2\to 0}\psi_{\zeta}(V_1,V_2,V_3,V_4) & = & \psi_{\zeta}(\sigma_3 V_3, \sigma_4 V_4),
\nonumber \\
\lim_{V_1,V_2\to \infty} \psi_{\zeta}(V_1,V_2,V_3,V_4) & = & T^{(1)} \psi_{\zeta}(V_3,V_4).
\label{3.17}
\end{eqnarray}
Unlike the $\sigma_i$ which are determined while solving the Dirac equation, the $T^{(i)}$ are already known from the single 
breather problem, see Eq.~(\ref{2.42}).

\begin{figure}[h]
\begin{center}
\epsfig{file=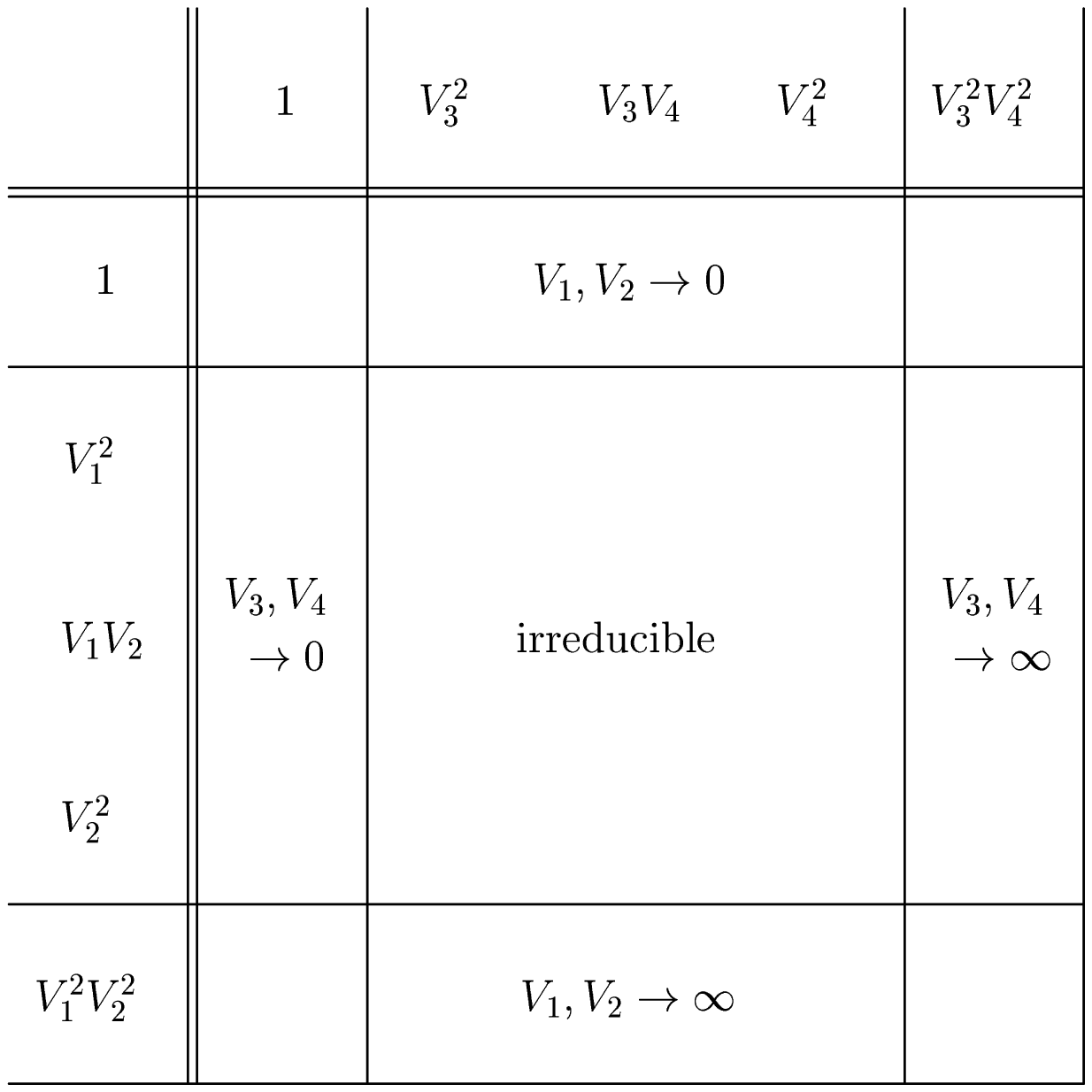,angle=270,width=6cm}
\caption{Square matrix into which the coefficients of ${\cal N},{\cal D},{\cal N}_1, {\cal N}_2$ for breather-breather scattering can be
naturally fitted. The innermost 3$\times$3 square contains the irreducible coefficients, all other coefficients are asymptotically reducible
as indicated in the figure.}
\label{Fig8}
\end{center}
\end{figure}

Inspection of the ansatz (\ref{3.4}) shows that these asymptotic relations relate all coefficients to single breather
coefficients and $\sigma_i$'s except for those multiplying $V_1^2V_3^2$, $V_1^2V_4^2$, $V_2^2 V_3^2$, $V_2^2 V_4^2$, $V_1 V_2 V_3^2$,
$V_1 V_2 V_4^2$, $V_1^2 V_3 V_4$, $V_2^2 V_3 V_4$, $V_1 V_2 V_3 V_4$. This reduces the number of parameters in $S$ and $\psi_{\zeta}$
from 2$\times$25 to 2$\times$9. This reduction is less effective than for baryons, where the corresponding reduction is from 
2$\times$9 to 2$\times$1 parameters, but nonetheless useful for solving the Dirac equation algebraically. These various numbers

for baryons and breathers have actually a simple interpretation, depicted in Fig.~\ref{Fig8}.
Since our ansatz (\ref{3.4}) has been derived by multiplying 2 polynomials for single breathers, the coefficients fit naturally into a square matrix. 
From Eq.~(\ref{3.17}), we see that the first and last row and the first and last column of this matrix are filled with reducible parameters, 
whereas the innermost 3$\times$3 square contains all irreducible coefficients. For baryons, the corresponding square is only 
of size 3$\times$3 and the irreducible inner part a single element. This explains the above counting of reducible and irreducible coefficients.

\subsection{Results for scalar potential and continuum spinors}\label{sect3b}
As explained in detail in Refs.~\cite{L8,L9}, the unknown coefficients can be determined algebraically by inserting the ansatz for $S$ and the 
spinors $\psi_{\zeta}$ into  the Dirac equation. Here, we immediately turn to the results for breather-breather scattering. 
We first  report on the complex scattering factors $\sigma_i$ describing how $V_i$ in one breather is affected by the collision 
with the other breather. General properties of the $\sigma_i$ have been given in Eqs.~(\ref{3.14}-\ref{3.16}).
We find that the $\sigma_i$ factorize as follows,
\begin{eqnarray}
\sigma_1 & = &  \theta_{13} \theta_{14}, \quad \sigma_2  = \theta_{23} \theta_{24},
\nonumber \\
\sigma_3 & = &  \theta_{31} \theta_{32}, \quad \sigma_4  = \theta_{41} \theta_{42}.
\label{3.18}
\end{eqnarray} 
The individual factors $\theta_{ij}$ may be interpreted as scattering amplitude of $V_i$ on $V_j$, where each $V_i$ stands for a
kink or antikink constituent of one of the breathers, and have the values
\begin{eqnarray}
\theta_{13} & = & \frac{1}{\theta_{31}} = \frac{\eta_1 Z_2-\eta_2 Z_1}{\eta_1 Z_2+\eta_2 Z_1},
\nonumber \\
\theta_{14} & = & \frac{1}{\theta_{41}} =\frac{\eta_1 + \eta_2 Z_1 Z_2}{\eta_1-\eta_2 Z_1 Z_2},
\nonumber \\
\theta_{23} & = & \frac{1}{\theta_{32}} =\frac{\eta_1 Z_1 Z_2 + \eta_2}{\eta_1 Z_1 Z_2 - \eta_2},
\nonumber \\
\theta_{24} & = & \frac{1}{\theta_{42}} =\frac{\eta_1 Z_1-\eta_2 Z_2}{\eta_1 Z_1 + \eta_2 Z_2}.
\label{3.19}
\end{eqnarray}
Just like in the baryon case, one can actually understand the results (\ref{3.18},\ref{3.19}) in simple terms.
First notice that the bound states can be identified via the singularities of the fermion transmission amplitudes,
i.e., the poles of $T^{(1)}$ and $T^{(2)}$, Eq.~(\ref{2.42}), in the complex $\zeta$-plane located at
\begin{eqnarray}
\zeta_1 & = & \frac{Z_1}{\eta_1}, \quad \zeta_2 = - \frac{1}{Z_1\eta_1},
\nonumber \\
\zeta_3 & = & \frac{Z_2}{\eta_2}, \quad \zeta_4 = - \frac{1}{Z_2 \eta_2}.
\label{3.20}
\end{eqnarray}
For a single breather, this has been discussed above, see Eqs.~(\ref{2.42a}-\ref{2.42c}).
Expressing the $\theta_{ij}$ through these complex $\zeta$-values yields the simple expression
\begin{equation}
\theta_{ij}  =   (\theta_{ji})^{-1} =  \frac{\zeta_j - \zeta_i}{\zeta_j+\zeta_i}, \quad  (i<j).
\label{3.21}
\end{equation}
This enables us to identify the scattering factors $\sigma_i$ with transmission amplitudes evaluated at complex $\zeta$ values,
i.e., for bound states. First we rewrite $T^{(i)}$ using the variables $\zeta_i$ rather than
$\eta_i, Z_i$, so that the pole structure is manifest,
\begin{eqnarray}
T^{(1)}(\zeta) & = & \frac{\zeta+\zeta_1}{\zeta - \zeta_1} \frac{\zeta +\zeta_2}{\zeta-\zeta_2},
\nonumber \\
T^{(2)}(\zeta) & = & \frac{\zeta+\zeta_3}{\zeta - \zeta_3} \frac{\zeta +\zeta_4}{\zeta-\zeta_4}.
\label{3,22}
\end{eqnarray}
A comparison with (\ref{3.18}) then shows that
\begin{eqnarray}
\sigma_1 & = & (T^{(2)}(\zeta_1))^{-1},
\nonumber \\
\sigma_2 & = &  (T^{(2)}(\zeta_2))^{-1},
\nonumber \\
\sigma_3 & = & T^{(1)} (\zeta_3),
\nonumber \\
\sigma_4 & = & T^{(1)} (\zeta_4).
\label{3.23}
\end{eqnarray}
Thus, like for baryons, one can relate the two-body scattering data to single breather input, namely the transmission amplitude
for bound states, evaluated at complex spectral parameters.
Notice also that the total fermion transmission amplitude for the two-breather system factorizes into $T^{(1)} T^{(2)}$ with the 
symmetric result 
\begin{equation}
T(\zeta) = \prod_{i=1}^4 \frac{\zeta+\zeta_i}{\zeta-\zeta_i}.
\label{3.24}
\end{equation}
Since all of these asymptotic quantities are independent of $Q_i$, they are actually identical in the breather and baryon cases (for the same $Z_i,
\eta_i$).

Equipped with the scattering factors $\theta_{ij}$, we can now present the results for all the coefficients in a concise form. Although the 
asymptotic conditions yield only the reducible coefficients along the periphery of the square shown in Fig.~\ref{Fig8}, 
it turns out that all coefficients except the center element ($i=j=3$) can be generated by a simple algorithm from single
breather input. Thus only the coefficients $a_{1234},b_{1234},c_{1234},d_{1234}$ are truly irreducible, exactly like in the baryon case.
Note that the reduction we are now talking about became only apparent after solving the problem and was not anticipated by us,
not being related to asymptotics. Nevertheless, it is very useful for presenting the results for a large number of coefficients in
a compact fashion.

Let us first explain how to generate all the coefficients except the ones multiplying $V_1V_2V_3V_4$. Using the matrix
scheme shown in Fig.~\ref{Fig8}, we introduce the 5$\times$5 square matrix
\begin{equation}
{\bf M}  =  \left( \begin{array}{ccccc} 
1    &   \theta_{31}^2 \theta_{32}^2     & \theta_{31} \theta_{32} \theta_{41} \theta_{42}   &  \theta_{41}^2 \theta_{42}^2  & \theta_{31}^2 \theta_{32}^2
\theta_{41}^2 \theta_{42}^2   \\
1    &  \theta_{32}^2    &  \theta_{32}\theta_{42}   & \theta_{42}^2  &  \theta_{32}^2 \theta_{42}^2  \\ 
1    &   \theta_{31} \theta_{32}    &   0   &  \theta_{41} \theta_{42}  &  \theta_{31} \theta_{32} \theta_{41} \theta_{42}  \\
1    &   \theta_{31}^2    &  \theta_{31} \theta_{41}   &  \theta_{41}^2  &   \theta_{31}^2 \theta_{41}^2  \\
1    &   1    &  1    & 1  & 1  
\end{array} \right).   
\label{MAT} 
\end{equation}
Consider the numerator  ${\cal N}$ of $S$
first, and write down the numerators ${\cal N}^{(1)}$ and ${\cal N}^{(2)}$ of $S$ for the individual breathers 1 and 2, cf. Eq.~(\ref{2.37}).
Convert the polynomials ${\cal N}^{(1,2)}$ 
into vectors according to their monomial content, using the same basis as in Fig.~\ref{Fig8},
\begin{equation}
{\bf U}_a^{(1)}  = \left( \begin{array}{c}    1 \\ a_{11}^{(1)} V_1^2 \\ a_{12}^{(1)} V_1 V_2 \\ a_{22}^{(1)} V_2^2 \\ a_{1122}^{(1)} V_1^2V_2^2
\end{array} \right), \quad {\bf U}_a^{(2)} = \left( \begin{array}{c}
1 \\ a_{11}^{(2)} V_3^2 \\ a_{12}^{(2)} V_3 V_4 \\ a_{22}^{(2)}V_4^2 \\ a_{1122}^{(2)} V_3^2V_4^2 \end{array} \right).
\label{3.27}
\end{equation}
All coefficients appearing here are single breather coefficients given in Eq.~(\ref{2.38}). Then, multiplying the matrix ${\bf M}$ with
 ${\bf U}_a^{(1)}$
from the left and ${\bf U}_a^{(2)}$ from the right yields ${\cal N}$ for breather-breather scattering, correct except for the
single coefficient $a_{1234}$,
\begin{equation}
{\cal N} = {\bf U}_a^{(1)} {\bf M} {\bf U}_a^{(2)}.
\label{3.28}
\end{equation}
The one missing, non-trivial coefficient $a_{1234}$ will be given explicitly below. 

Somewhat miraculously, the 
same procedure works for the $b$-coefficients of the denominator ${\cal D}$ of $S$, as well as for the $c,d$-coefficients of the 
numerators ${\cal N}_{1,2}$ of the  continuum spinors. The matrix ${\bf M}$ is always the same, so that this 
representation of the results is very economical indeed. In all cases, the vectors ${\bf U}$ can be constructed from
the single breather results in a fashion analogous to what was done in Eq.~(\ref{3.27}).
This boils down to replacing all the $a^{(i)}$ coefficients by the corresponding $b^{(i)},c^{(i)},d^{(i)}$, see Eqs.~(\ref{2.38},\ref{2.41}).
The resulting vectors will be denoted by ${\bf U}_b^{(i)}, {\bf U}_c^{(i)}, {\bf U}_d^{(i)}$ accordingly, and we get
\begin{eqnarray}
 {\cal D} & = &  {\bf U}_b^{(1)} {\bf M} {\bf U}_b^{(2)},
 \nonumber \\
 {\cal N}_1 & = &  {\bf U}_c^{(1)} {\bf M} {\bf U}_c^{(2)},
 \nonumber \\
 {\cal N}_2 & = &  {\bf U}_d^{(1)} {\bf M} {\bf U}_d^{(2)}.
\label{3.28a}
\end{eqnarray}
These relations yield the correct coefficients except for $b_{1234},c_{1234},d_{1234}$.

This completes the presentation of all the coefficients entering the scalar potential and the continuum spinors for breather-breather
scattering, except for those multiplying $V_1V_2V_3V_4$. The 4 missing coefficients $a_{1234}, b_{1234}, c_{1234}, d_{1234}$
are the most complicated ones, not related to single breather input in any obvious way. 
Actually, we have found that they are proportional to the irreducible baryon-baryon coefficients $a_{12}^{\rm B},b_{12}^{\rm B},c_{12}^{\rm B},
d_{12}^{\rm B}$ of Ref.~\cite{L8} with a simple proportionality factor depending only on the $Q_i$,
\begin{equation}
a_{1234}  =  \frac{2Q_1}{Q_1^2+1}\frac{2Q_2}{Q_2^2+1}  a_{12}^{\rm B} .
\label{3.29}
\end{equation}
This equation remains valid if we replace $a$ by $b,c,d$ on both sides, with exactly the same proportionality factor. 
The irreducible baryon coefficients have been spelled out in Ref.~\cite{L8} and are somewhat lengthy.
To keep the present paper self-contained, we list them here once again, using the more structured form given in Ref.~\cite{L9},
\begin{eqnarray}
a_{12}^{\rm B} & = & \frac{a_1^1 a_1^2}{d_{12}} ( \rho_1 \rho_2 + B_{12} ),
\nonumber \\
b_{12}^{\rm B} & = & \frac{b_1^1 b_1^2}{d_{12}}  B_{12},
\nonumber \\
c_{12}^{\rm B} & = & \frac{c_1^1 c_1^2}{d_{12}} ( \mu_1\mu_2 + B_{12} ),
\nonumber \\
d_{12}^{\rm B} & = & \frac{d_1^1 d_1^2}{d_{12}} ( \nu_1\nu_2 + B_{12} ),
\label{3.30}
\end{eqnarray}
with one-baryon coefficients
\begin{eqnarray}
a_1^i & = & - \frac{2(Z_i^4+1)}{Z_i(Z_i^2+1)},
\nonumber \\
b_1^i & = & -  \frac{4 Z_i}{Z_i^2+1},
\nonumber \\
c_1^i & = & \frac{2(Z_i^4+1-2 Z_i^2 \zeta^2 \eta_i^2)}{(Z_i^2+1)( \zeta \eta_i -Z_i)(\zeta \eta_i Z_i +1)},
\nonumber \\
d_1^{i} & = & \frac{2(2 Z_i^2- \zeta^2 \eta_i^2 (Z_i^4+1))}{(Z_i^2+1)( \zeta \eta_i -Z_i)( \zeta \eta_i Z_i+1)},
\nonumber \\
\rho_i & = & \frac{Z_i^4-1}{Z_i^4+1},
\nonumber \\
\mu_i & = & \frac{Z_i^4-1}{Z_i^4+1-2 Z_i^2 \zeta^2 \eta_i^2},
\nonumber \\
\nu_i & = & \frac{(Z_i^4-1)\eta_i^2 \zeta^2}{2 Z_i^2 - \zeta^2 \eta_i^2 (Z_i^4+1)},
\label{3.31}
\end{eqnarray}
and two-baryon coefficients
\begin{eqnarray}
d_{12} & = & \frac{2(\eta_1Z_1-\eta_2Z_2)(\eta_1Z_2-\eta_2Z_1)(\eta_1Z_1Z_2+\eta_2)(\eta_2Z_1Z_2+\eta_1)}
{\eta_1^2\eta_2^2(Z_1^4-1)(Z_2^4-1)},
\nonumber \\
B_{12} & = & \frac{2 Z_1^2 Z_2^2 (\eta_1^4+\eta_2^4) - \eta_1^2\eta_2^2 (Z_1^4+1)(Z_2^4+1)}{\eta_1^2\eta_2^2(Z_1^4-1)(Z_2^4-1)}.
\label{3.32}
\end{eqnarray}
This close relation between breather-breather and baryon-baryon scattering was unexpected, and we do not have a simple
explanation for the proportionality (\ref{3.29}).

Summarizing this subsection, there seems to be no major additional complication when going from baryon-baryon to 
breather-breather scattering, in spite of the increased number of parameters and the more complicated dynamical processes which
are now described. Our strategy of solving the Dirac equation via the ansatz method works equally well in both cases.
When we tried to present our results for the large number of coefficients in the breather case in the most compact fashion,
we discovered surprisingly close connections between breather-breather scattering and individual
breathers (for reducible coefficients), but also between breathers and baryons (for irreducible coefficients). 
These insights may give us hints as how to simplify and possibly generalize this whole calculation in the future, perhaps
by exploiting more efficiently the integrability of the GN model.

\subsection{Results for bound state spinors}\label{sect3c}
The bound state spinors can be found either by ansatz and the solution of the Dirac equation, or by analytic continuation from the
continuum spinors. In both cases, one has to take linear combinations which satisfy the orthogonality condition and the additional 
conditions on the scalar density discussed for the single breather bound states, Eqs.~(\ref{2.43},\ref{2.44}).
There are four bound states, two for each breather. The relevant ansatz has already been given in Sec.~{\ref{sect3a}.
We will again try to present the $e,f$-coefficients for states 1 and 3 in the most efficient way, looking for factorization of our results. 
The states 2
and 4 can then be obtained for free, using charge conjugation, Eq.~(\ref{3.12}). 

To present the results for the spinor of bound state 1 in a compact way, we introduce the following vectors,
\begin{equation}
{\bf W}_{1}^{\pm} = \left( \begin{array}{c} 0 \\ 1-Q_1 \\ 0 \\ 0 \\ \pm (1+Q_1)V_2^2 \end{array} \right), \quad
{\bf W}_{2}^{\pm} = \left( \begin{array}{c} 0 \\ 0 \\ 0 \\ 1+Q_1 \\ \pm (1-Q_1)V_1^2 \end{array} \right).
\label{3.32a}
\end{equation}
Then the numerators entering Eq.~(\ref{3.7}) can be represented as
\begin{eqnarray}
{\cal N}_1^{(1)} & = & Z_1 V_1 {\bf W}_{1}^{+} {\bf M} \left. {\bf U}_c^{(2)} \right|_{\zeta=\zeta_1}+ V_2 {\bf W}_{2}^{-} {\bf M} \left. \bf{U}_c^{(2)} \right|_{\zeta=\zeta_2}
\nonumber  \\
{\cal N}_2^{(1)} & = & V_1 {\bf W}_{1}^{-} {\bf M} \left. {\bf U}_d^{(2)} \right|_{\zeta=\zeta_1} - Z_1V_2 {\bf W}_{2}^{+} {\bf M} \left. \bf{U}_d^{(2)} \right|_{\zeta=\zeta_2}
\label{3.32b}
\end{eqnarray}
The normalization factor for the  spinor of bound state 1 can be inferred from the single breather at times where the two breathers 
are well separated,
\begin{equation}
C^{(1)} = C_0^{(1)}
\label{3.40}
\end{equation}
with $C_0^{(1)}$ from Eq.~(\ref{2.47}).

Similarly, the results for the spinor of bound state 3 require the vectors
\begin{equation}
{\bf W}_{3}^{\pm} = \left( \begin{array}{c} 0 \\ 1-Q_2 \\ 0 \\ 0 \\ \pm (1+Q_2)V_4^2 \end{array} \right), \quad
{\bf W}_{4}^{\pm} = \left( \begin{array}{c} 0 \\ 0 \\ 0 \\ 1+Q_2 \\ \pm (1-Q_2)V_3^2 \end{array} \right).
\label{3.40a}
\end{equation}
The numerators in Eq.~(\ref{3.10}) are then found to be
\begin{eqnarray}
{\cal N}_1^{(3)} & = & \theta_{13} \theta_{23} Z_2 V_3  {\bf W}_{3}^{+} {\bf M}^T \left. {\bf U}_c^{(1)} \right|_{\zeta=\zeta_3} 
+ \theta_{14} \theta_{24} V_4  {\bf W}_{4}^{-} {\bf M}^T \left. \bf{U}_c^{(1)} \right|_{\zeta=\zeta_4}
\nonumber  \\
{\cal N}_2^{(3)} & = &  \theta_{13} \theta_{23} V_3  {\bf W}_{3}^{-} {\bf M}^T \left. {\bf U}_d^{(1)} \right|_{\zeta=\zeta_3}
 -  \theta_{14} \theta_{24} Z_2V_4 {\bf W}_{4}^{+} {\bf M}^T \left. \bf{U}_d^{(1)} \right|_{\zeta=\zeta_4}
\label{3.40b}
\end{eqnarray}
The normalization factor for the spinors of bound state 3 is
\begin{equation}
C^{(3)} =  C_0^{(2)}
\label{3.46}
\end{equation}
with $C_0^{(2)}$ from Eq.~(\ref{2.47}). The extra factors $\theta_{ij}$ in (\ref{3.40b}) as compared to (\ref{3.32b}) must be a consequence of the definition
(\ref{3.21}) which distinguishes between $i>j$ and $i<j$.

\subsection{Self-consistency and fermion density}\label{sect3d}
The ansatz method yields a time-dependent, transparent scalar potential for the Dirac equation, with the correct
boundary conditions for breather-breather scattering. We still have to verify self-consistency so as to be sure
that we have found a TDHF solution of the GN model. The way self-consistency arises always follows the same pattern.
The scalar density for the continuum spinors can be decomposed into
\begin{eqnarray}
\bar{\psi}_{\zeta} \psi_{\zeta} & = &  (\bar{\psi}_{\zeta} \psi_{\zeta})_1 +  (\bar{\psi}_{\zeta} \psi_{\zeta})_2,
\nonumber \\
(\bar{\psi}_{\zeta} \psi_{\zeta})_1 & = & - \frac{2\zeta}{\zeta^2+1} S.
\label{3.47}
\end{eqnarray}
The first term gives self-consistency as usual, whereas the 2nd term can be represented as linear combination of the 
scalar densities of bound states, 
\begin{equation}
(\bar{\psi}_{\zeta} \psi_{\zeta})_2 = \alpha^{(1)} \bar{\psi}^{(1)} \psi^{(1)} + \alpha^{(3)} \bar{\psi}^{(3)} \psi^{(3)}.
\label{3.48}
\end{equation}
Upon integrating over $d\zeta$ and making use of the self-consistency conditions for the individual breathers with occupation
fractions $\nu^{(1,2)}$, we find analytically
\begin{eqnarray}
\int_0^{\infty} \frac{d\zeta}{2\pi} \left( \frac{\zeta^2+1}{2\zeta^2} \right) \alpha^{(1)} & = & - (1- \nu^{(1)}),
\nonumber \\
\int_0^{\infty} \frac{d\zeta}{2\pi} \left( \frac{\zeta^2+1}{2\zeta^2} \right) \alpha^{(3)} & = & -(1- \nu^{(2)}).
\label{3.49}
\end{eqnarray}
This is equal and opposite to the contributions from the discrete states and therefore cancelled exactly if we sum over
continuum and bound states.
Like baryon-baryon scattering, breather-breather scattering is a type III TDHF solution.   

Consider the fermion density next. We expect that the situation is the same as for baryon-baryon scattering, i.e., the total, vacuum subtracted
fermion density can be expressed in terms of the bound state densities $\rho^{(1)}=\rho^{(2)}$ and $\rho^{(3)}=\rho^{(4)}$ as
\begin{equation}
\rho_{\rm tot} = N \left( \nu^{(1)} \rho^{(1)}+ \nu^{(2)} \rho^{(3)} \right).
\label{3.50}
\end{equation}
The total fermion number is
\begin{equation}
N_f = N \left( \nu^{(1)} + \nu^{(2)} \right).
\label{3.51}
\end{equation}
Like for the single breather, the density for a bound state can most conveniently be represented in terms of a pseudoscalar field $P$, 
see Eqs.~(\ref{2.48b}-\ref{2.48e1}). The result for bound state 1 in breather-breather scattering has the following simple form,
\begin{eqnarray}
\rho^{(1)} & = & 
\psi^{(1)\dagger} \psi^{(1)}  =  \frac{\partial}{\partial x} \frac{{\cal T}_1}{\cal D},
\nonumber \\
{\cal T}_1 & = & \frac{1}{2} {\bf U}_b^{(1)} {\rm diag}(-1,0,0,0,1) {\bf M} {\bf U}_b^{(2)}.
\label{3.51a}
\end{eqnarray}
Similarly, we find for bound state 3 
\begin{eqnarray}
\rho^{(3)} & = & 
\psi^{(3)\dagger} \psi^{(3)}  =  \frac{\partial}{\partial x} \frac{{\cal T}_3}{\cal D},
\nonumber \\
{\cal T}_3 & = & \frac{1}{2} {\bf U}_b^{(1)}{\bf M}\, {\rm diag}(-1,0,0,0,1) {\bf U}_b^{(2)}.
\label{3.51b}
\end{eqnarray}
Identity (\ref{3.50}) is then best verified numerically. This close relation between the total fermion density and the
density from the bound states is the 
reason why we discussed the bound states in some detail. Eqs.~(\ref{3.50}-\ref{3.51b}) will be needed in the following to compute the fermion density in
breather-breather scattering analytically.

\subsection{Illustrative results}\label{sect3e}
We have presented the full, analytical solution of breather-breather scattering in the GN model, exact in the large $N$ limit.  
The asymptotic scattering information displays complete factorization, and can in fact be predicted on the basis of the single breather.
This is a manifestation of the integrability of the GN model. The (reflectionless) fermion-double breather transmission amplitude is simply
the product of two independent fermion-breather amplitudes, Eq.~(\ref{3.24}). It has poles in the complex plane of the spectral parameter
$\zeta$ corresponding to four bound states, two per breather. The asymptotic breather-breather 
scattering information is encoded in the complex numbers $\sigma_i$ which multiply the basis exponentials $V_i$ in the outgoing channel, 
see Figs.~\ref{Fig6} and \ref{Fig7}. As usual for solitons, they lead to a time delay, but also to a change in the phase of the breather
oscillation. They can be inferred by evaluating fermion-breather transmission amplitudes for one breather
at the complex spectral parameters $\zeta_k$ corresponding to a bound state of the other breather, see Sect.~\ref{sect3b}.
This generalizes similar findings for baryon-baryon scattering in Ref.~\cite{L8} to breather-breather scattering. 

The TDHF solution does not only yield the $S$-matrix, but also the full time evolution of the fermion wave functions in the whole space.
Since the mean field is a classical quantity, we can illustrate breather-breather scattering in the same way as in
classical, nonlinear systems like the sine-Gordon model. There is nothing wrong in prescribing simultaneously the initial position and 
velocity of the breathers, as they are not subjected to the uncertainty principle. The novel aspect of these dynamically generated breathers as compared to 
classical bosonic theories are the fermions. Fermions have been treated in a fully quantum mechanical manner, using the TDHF approach.
They populate the breathers, and their dynamics can be followed in detail by monitoring the fermion density. Although the polarization
of the Dirac sea is fully taken into account, we have seen that the total fermion density can be computed using only bound state spinors
and occupation fractions. Since the self-consistent TDHF potential in the GN model is always reflectionless, one expects that the kinks and 
antikinks making up the breather will repel each other, whereas the fermions should always move forwards. This behavior has already
been observed in the case of baryon scattering. 

Since this paper is somewhat technical, let us first summarize how a full scattering calculation can be done with CA systems like Maple
or Mathematica. This should help the reader to navigate through our rather complicated set of equations.
One first has to set up the matrix ${\bf M}$, Eq.~(\ref{MAT}), the vectors ${\bf U}^{(1,2)}_{a,b,c,d}$, 
Eq.~(\ref{3.27}) and the vectors ${\bf W}_i^{\pm}$, Eqs.~(\ref{3.32a},\ref{3.40a}). The required coefficients are the single breather
 coefficients which can be found in (\ref{2.38}) for $a,b$-coefficients and in (\ref{2.41}) for $c,d$-coefficients.
Eqs.~(\ref{3.28},\ref{3.28a},\ref{3.32b},\ref{3.40b}) can then be used to generate conveniently
the basic polynomials ${\cal N},{\cal D}$ entering the TDHF potential $S$, ${\cal N}_{1,2}$ of the continuum spinors and ${\cal N}_{1,2}^{(1.3)}$
of the bound state spinors for bound states 1 and 3. This construction misses the irreducible terms which have to be added
by hand, using the $a,b,c,d$-coefficients from (\ref{3.29}-\ref{3.32}). The normalization factors for the bound states are given in (\ref{2.47},\ref{3.40},\ref{3.46}).
Together with the basis exponentials $V_k$, Eq.~(\ref{3.1}), these ingredients enable us to construct the mean field $S$
from Eqs.~(\ref{3.2}-\ref{3.4}),
the continuum spinor $\psi_{\zeta}$ from (\ref{3.5}), the bound state spinor $\psi^{(1)}$ from (\ref{3.7}-\ref{3.8}) and $\psi^{(3)}$
from (\ref{3.10}-\ref{3.11}). The fermion density can be readily evaluated using Eqs.~(\ref{3.50}-\ref{3.51b}).
This procedure may sound a bit complicated, but we remind the reader that we have to communicate 180 coefficients, each 
consisting of several factors. We have not found any simpler ways of making our results available. 

One can actually apply these results to various related problems. Choosing the two breather velocities differently, one
deals with the breather-breather scattering problem in an arbitrary Lorentz frame. Choosing the same velocity also yields a solution, 
a (marginal) bound state of two breathers
with mass equal to the sum of the two constituents. The distance between the breathers can be arbitrarily chosen, like for baryons and 
their composites
\cite{L3,L8}. If we set one of the $Q$-parameters equal to $-1$, the corresponding breather reduces to a baryon and we can handle 
breather-baryon scattering 
and bound states. If we then tune the $Z$ parameter of the baryon so that it becomes a widely separated kink-antikink pair ($Z\to i$ or $\epsilon \to 
\infty$), we can send one of the kinks to spatial infinity and account for breather-kink scattering and bound states. This encompasses
in particular
the analog of the breather-kink bound state that has been called a ``wobble" in sine-Gordon theory \cite{L17,L18}, a topologically 
non-trivial breather. 
Finally, if we set both $Q$'s equal to $-1$ we recover baryon-baryon, baryon-kink and kink-kink dynamics. This is nothing new but can serve
as a useful test of the algebra.

Since the results are in closed anlaytical form and their implementation is straightforward using CA, there is no point in showing many 
concrete examples here. The best way of illustrating these time dependent results is anyway by means of animations of the scalar
potential and the femion density. This is easy on a computer using Maple or Mathematica, see \cite{anim} for animations
of the examples discussed in the present work. 

In the first example we choose the breather with parameters $\epsilon=2, \lambda=2.8$ as incident from the left and the breather 
with $\epsilon=2.0, \lambda=1.1$ of Fig.~\ref{Fig2} as incident
from the right with equal and opposite velocities ($\eta_1 =\eta_2^{-1} = 2$). Fig.~\ref{Fig9} shows the time evolution of the scalar potential. One 
recognizes
the oscillations of the incoming and outgoing breathers as well as the collision region. Fig.~\ref{Fig10} gives a view at the fermion densities of these 
colliding breathers. It is hard to understand what exactly is happening from these plots.
We therefore give a few snapshots of the scalar potential and the fermion density  
during the collision in Fig.~11.  The fat curves are $S$, the thin solid curves are the fermion density from bound state 1 and the dashed 
thin curve the fermion density from bound state 3. Since these are separately conserved, it makes sense to split up the total fermion density
in this manner. This enables us to keep track of the individual lumps of fermions which indeed cross each other, each one 
moving only in the forward direction.
\begin{figure}[h]
\begin{center}
\epsfig{file=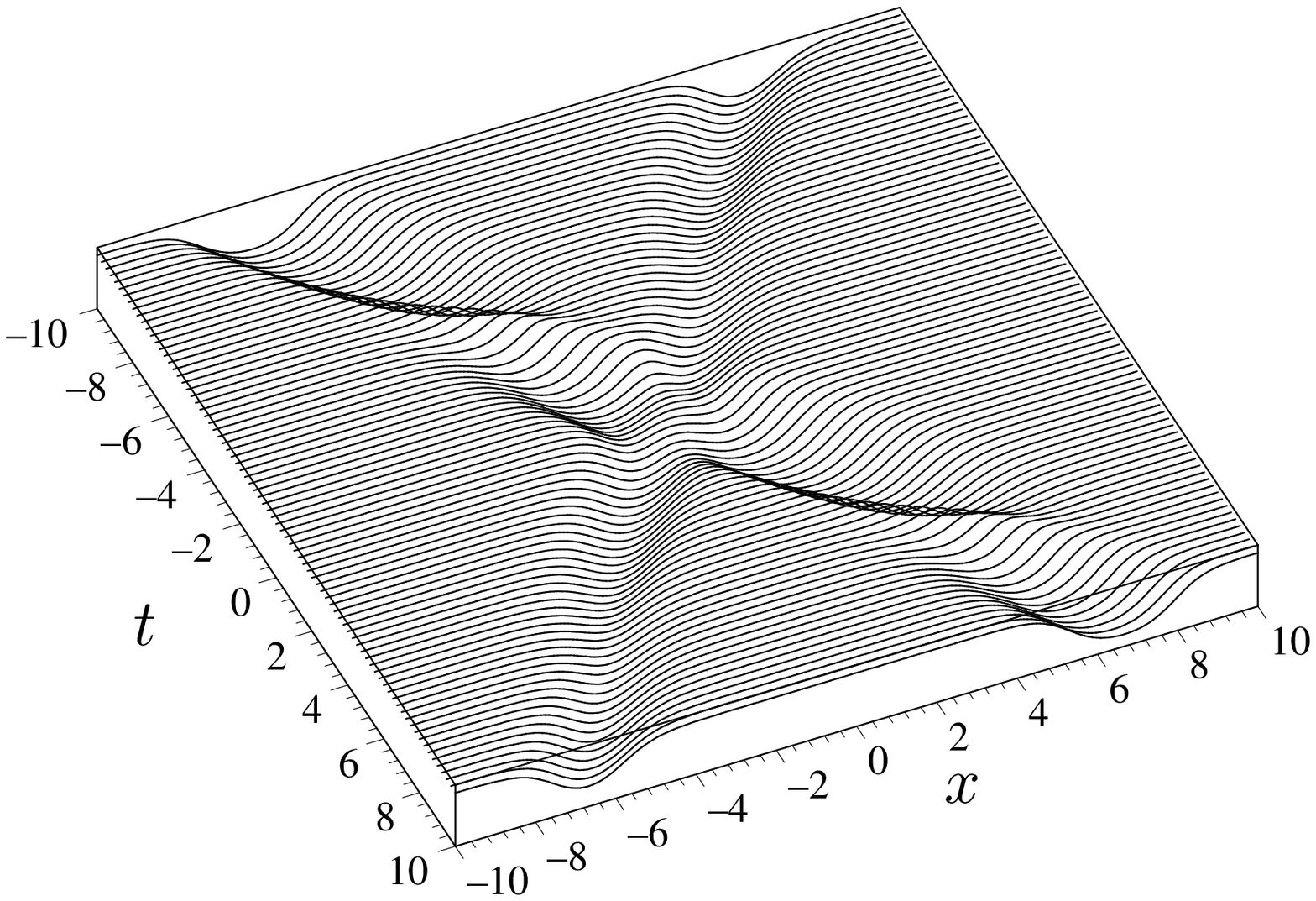,angle=270,width=12cm}
\caption{Time evolution of scattering of breathers with parameters $\epsilon=2, \lambda=2.8, \nu=0.1735$ (breather 1)and $\epsilon=2, \lambda=1.1, \nu=0.6753$ (breather 2),
during collision with $\eta_1=1/\eta_2=2$. See \cite{anim} for animations.}
\label{Fig9}
\end{center}
\end{figure}
\begin{figure}[h]
\begin{center}
\epsfig{file=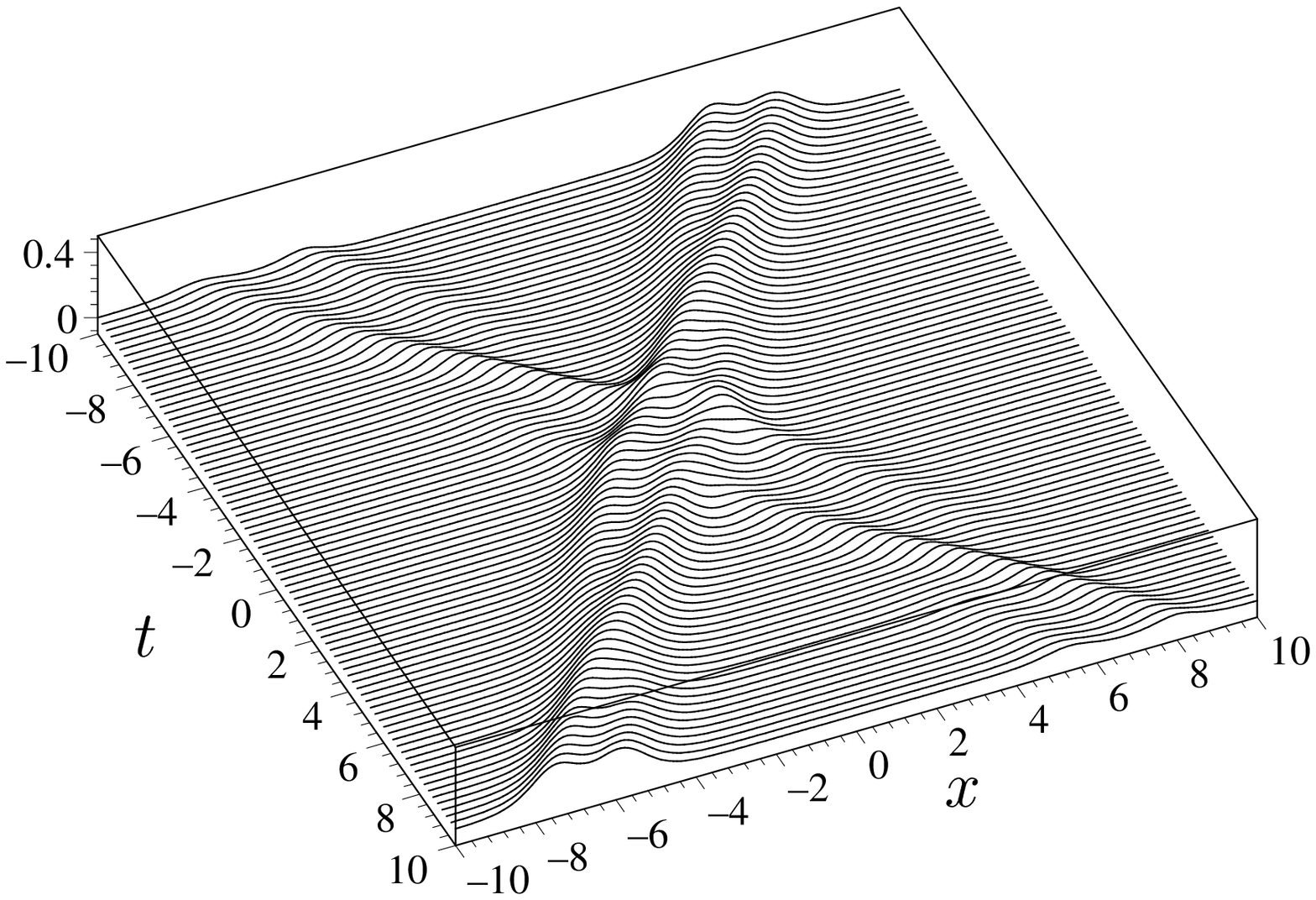,angle=270,width=12cm}
\caption{Time dependence of fermion density for the collision process of Fig.~\ref{Fig9}. See \cite{anim} for animations.}
\label{Fig10}
\end{center}
\end{figure}
\begin{figure}[h]
\begin{center}
\epsfig{file=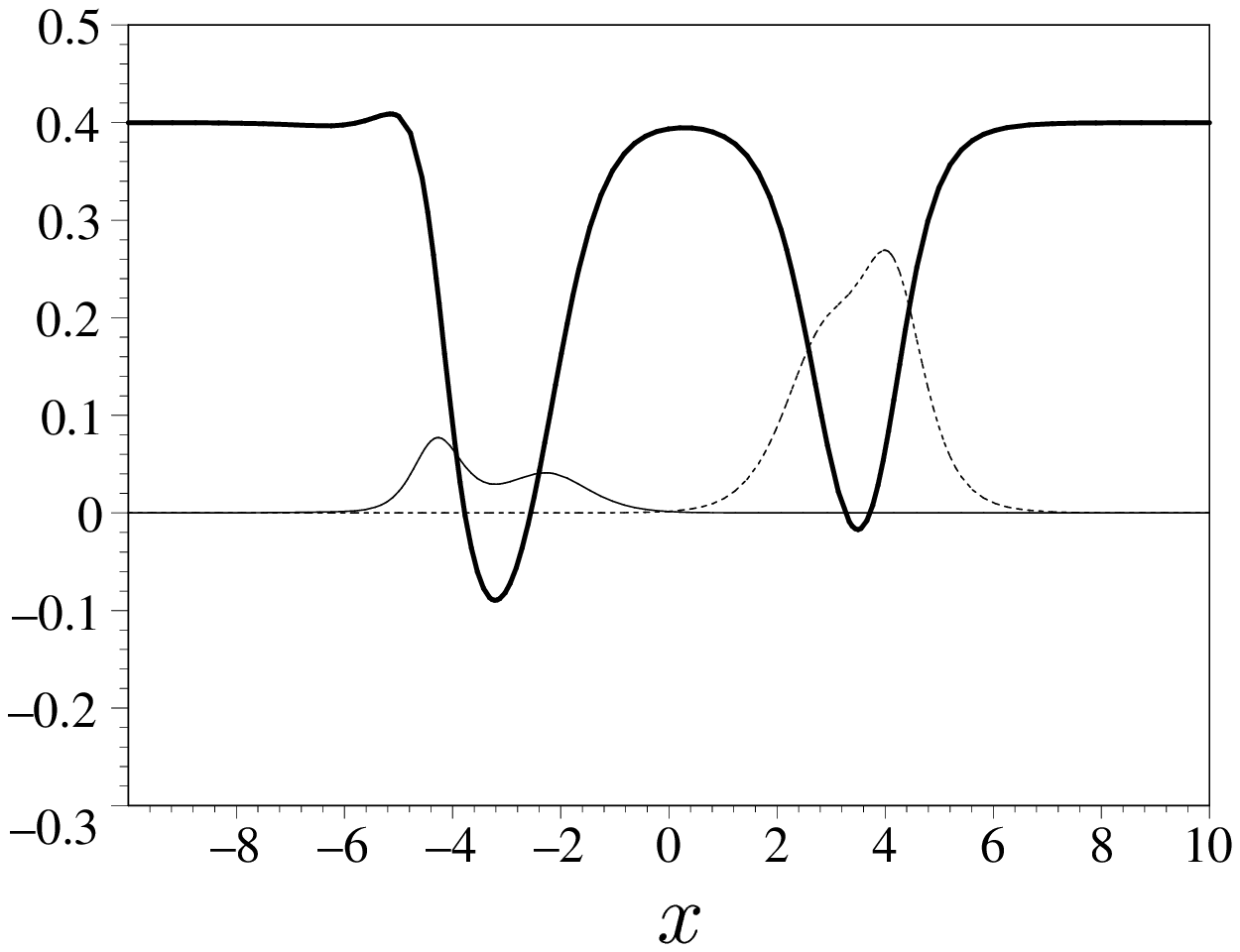,angle=270,width=5cm}\epsfig{file=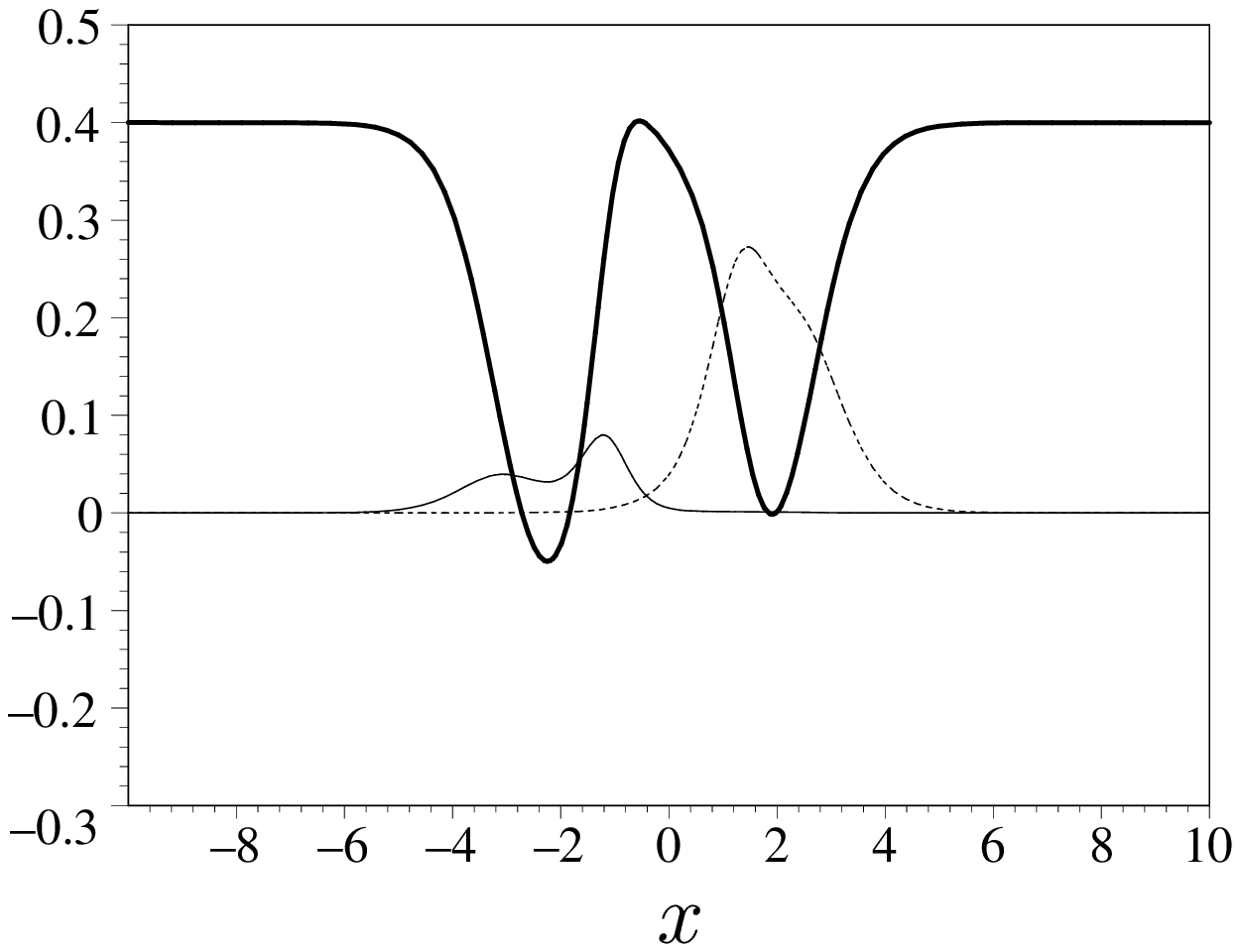,angle=270,width=5cm}
\epsfig{file=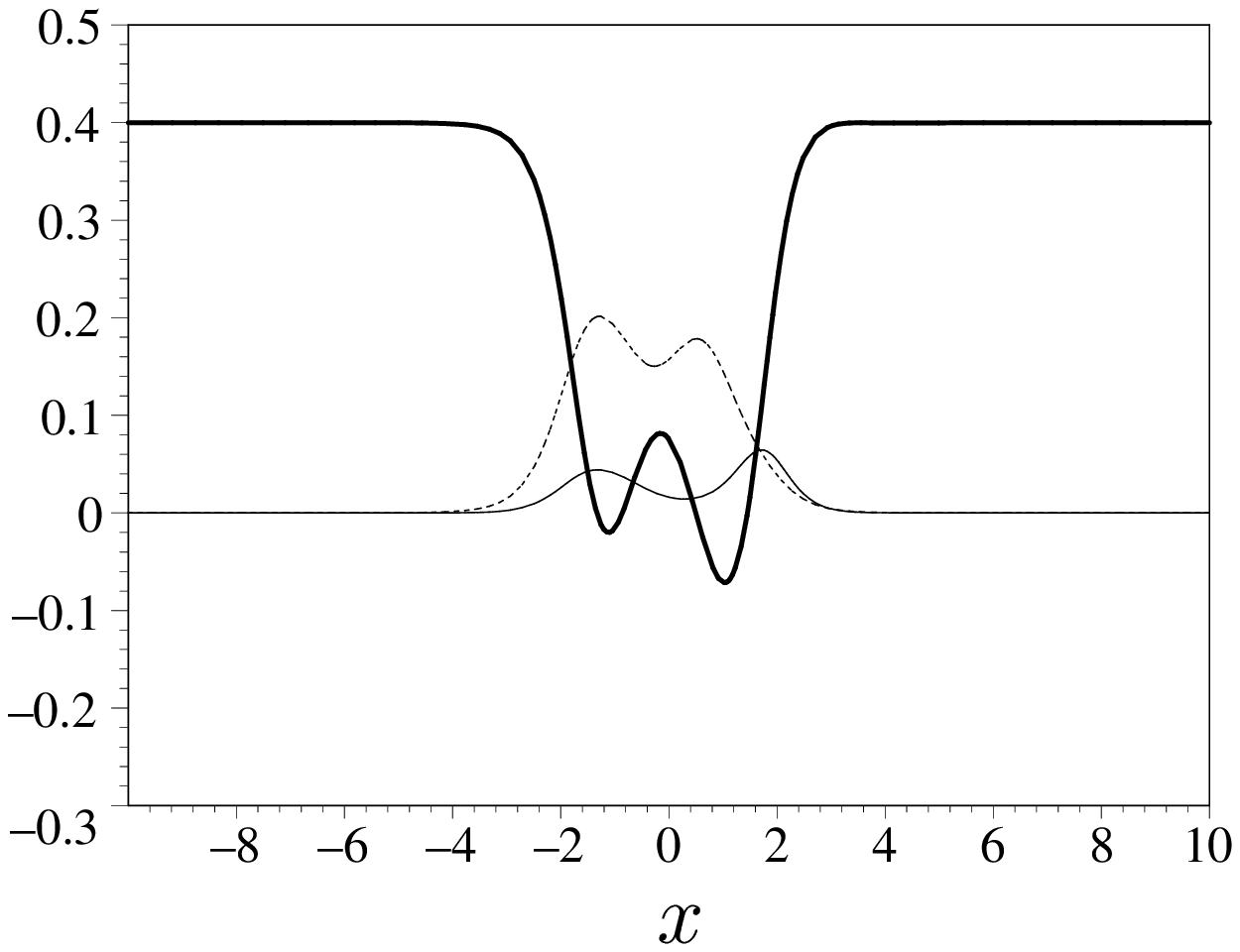,angle=270,width=5cm}\\
\epsfig{file=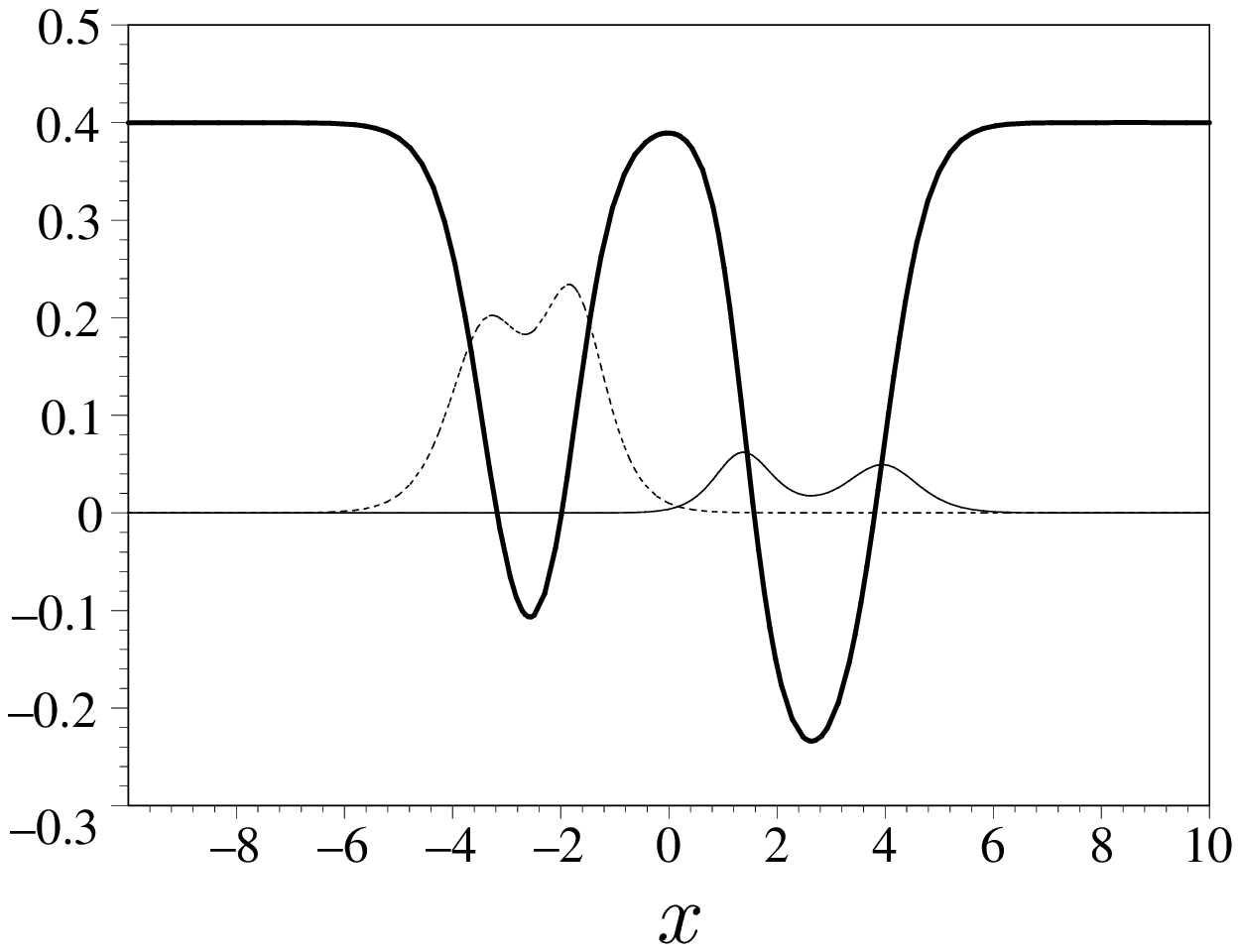,angle=270,width=5cm}\epsfig{file=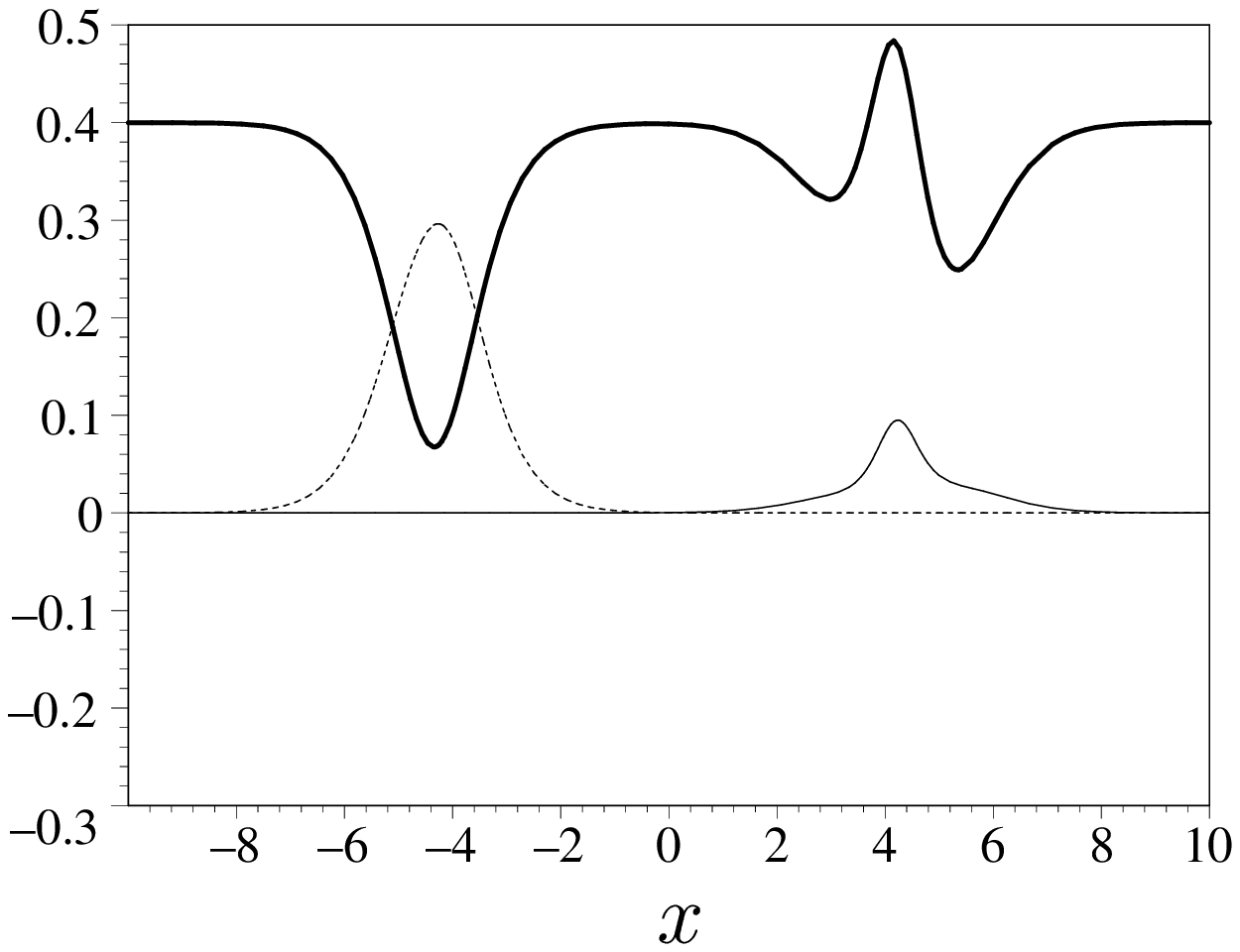,angle=270,width=5cm}
\caption{Snapshots of the collision process of Figs.~\ref{Fig9} and \ref{Fig10} at times -6,-3,0,3,6. Fat line: $0.4\times S$, solid thin line: $\rho^{(1)}$,
dashed thin line: $\rho^{(3)}$. See main text.}
\label{Fig11}
\end{center}
\end{figure}

In our 2nd example we collide two breathers with a more pronounced kink-antikink structure with parameters $\epsilon=700, \lambda= \sqrt{1+\epsilon^2}$ and $\epsilon=700,
\lambda=400$. The boost parameter $\eta_1=1/\eta_2=1.005$ now corresponds to a low relative velocity so as to be able to illustrate the ``breathing" 
and the interaction dynamics on a similar time scale. Fig.~{\ref{Fig12} shows the scalar potential, Fig.~\ref{Fig13} the total fermion density during the collision.
Here, complicated things happen with the fermions which seem to disappear and reappear elsewhere in an almost discontinuous
fashion. To further explore what is going on during the collision, in Fig.~\ref{Fig14} we present again a sequence of snapshots. 
In the first and last frame, the two fermion bound states are attached to their ``home" breather, as expected asymptotically.
Inbetween however, we see that the fermions hop from one kink or antikink to the next, spending most of the time
near the nodes of the potential. Kinks and antikinks approach each other up to some minimal distance where they are reflected.
At this point of closest approach the fermions tunnel through, as can be seen in an animation by the disappearance of one peak 
in the density and the simultaneous reappearance of another peak elsewhere. These tunneling processes are behind the 
apparently discontinuous behavior of the density in Fig.~\ref{Fig13}. In the collision region, the two individual bound state densities
of states 1 and 3 do not belong to any particular breather. In the 3rd frame of Fig.~\ref{Fig14}, bound state 1 is located at the outer
kink-antikink pair, bound state 2 at the inner kink-antikink pair. This is the (temporal) midpoint of the collision. 
In the neighboring frames the fermions are attached to 2 or even 3 different kinks and antikinks.
 This figure allows us to follow the hopping mechanism from kink to kink in detail.
Similar hopping processes were also observed in kink-antikink scattering processes \cite{L6,L7}. Clearly,
they are intimately connected to the fact that the self-consistent potential is transparent.

\begin{figure}[h]
\begin{center}
\epsfig{file=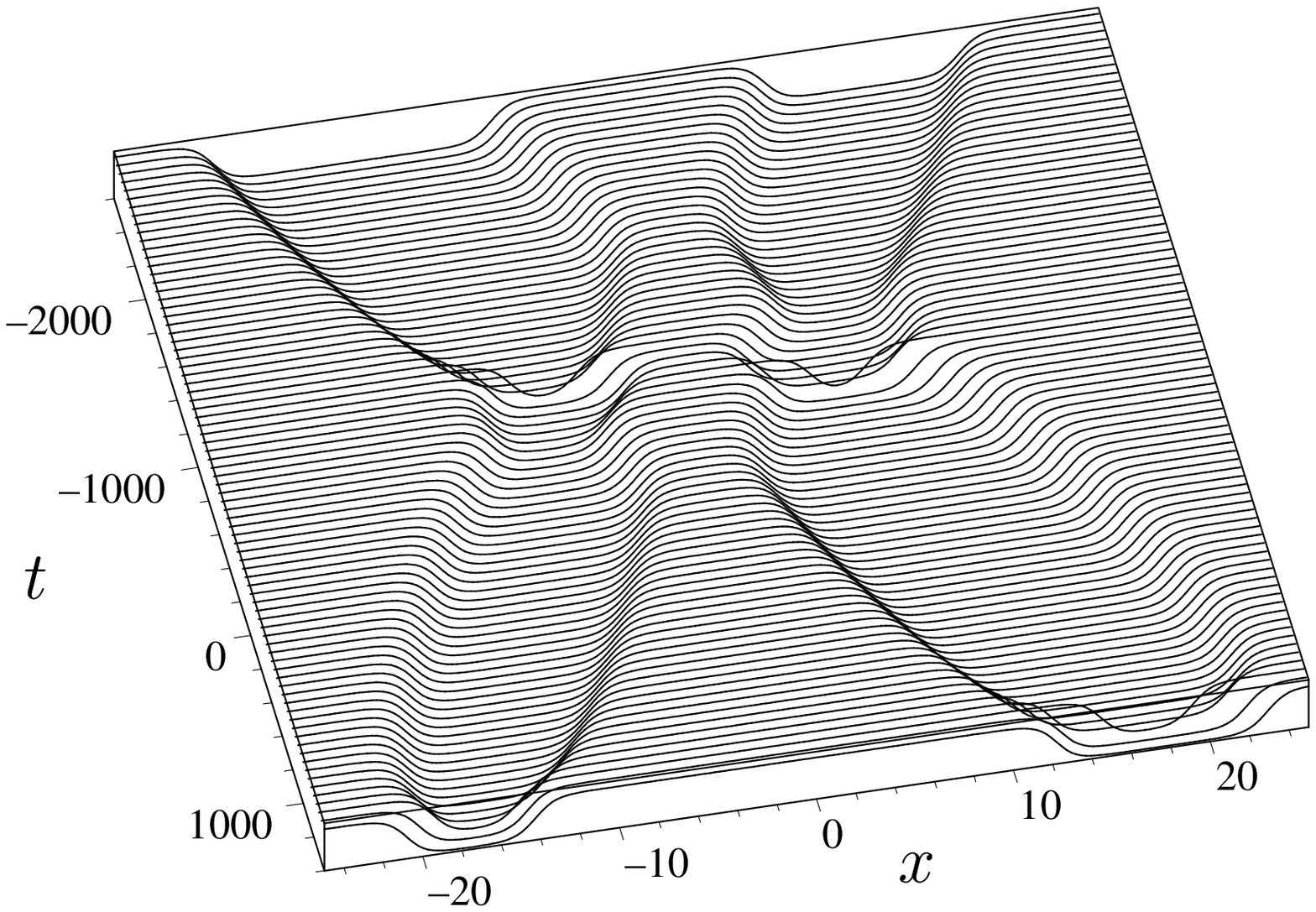,angle=270,width=12cm}
\caption{Time evolution of scattering of breathers with parameters $\epsilon=700, \lambda=\sqrt{1+\epsilon^2}, \nu=0.3634$ (breather 1) and
$\epsilon=700, \lambda=400, \nu=0.6362$ (breather 2), during collision with $\eta_1=1 / \eta_2=1.005$. See \cite{anim} for animations.}
\label{Fig12}
\end{center}
\end{figure}
\begin{figure}[h]
\begin{center}
\epsfig{file=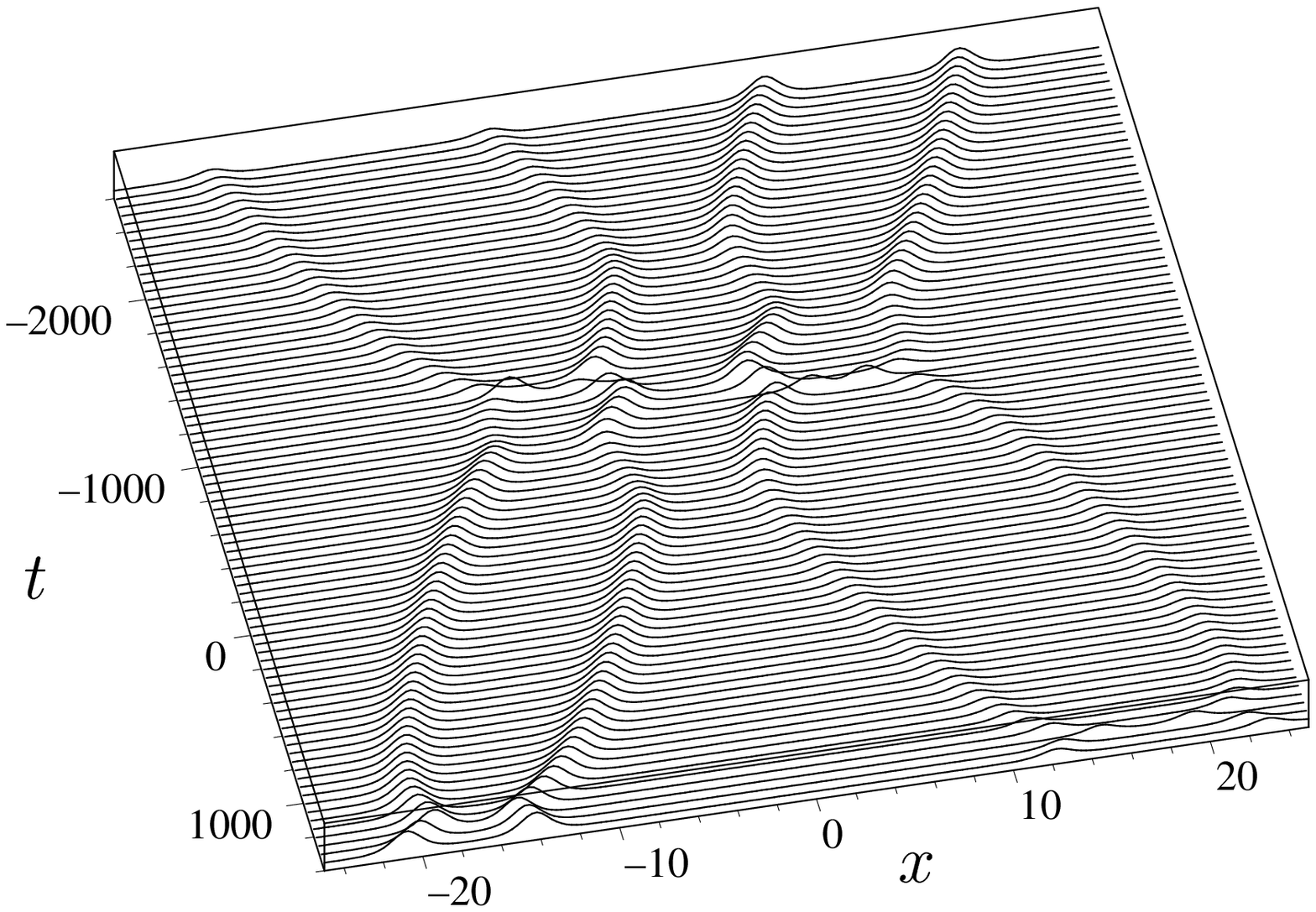,angle=270,width=12cm}
\caption{Time dependence of fermion density for the collision process of Fig.~\ref{Fig12}. See \cite{anim} for animations.}
\label{Fig13}
\end{center}
\end{figure}
\begin{figure}[h]
\begin{center}
\epsfig{file=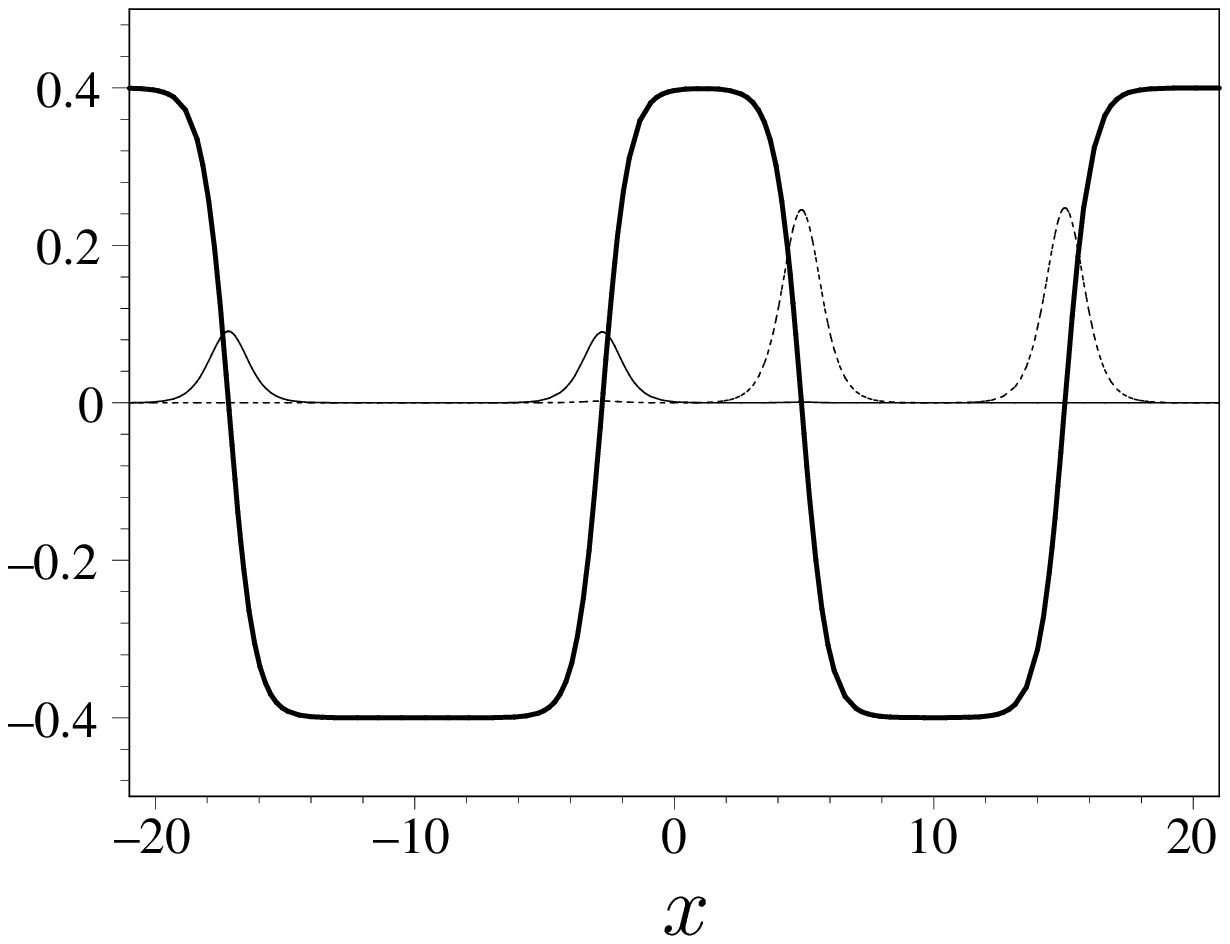,angle=270,width=5cm}\epsfig{file=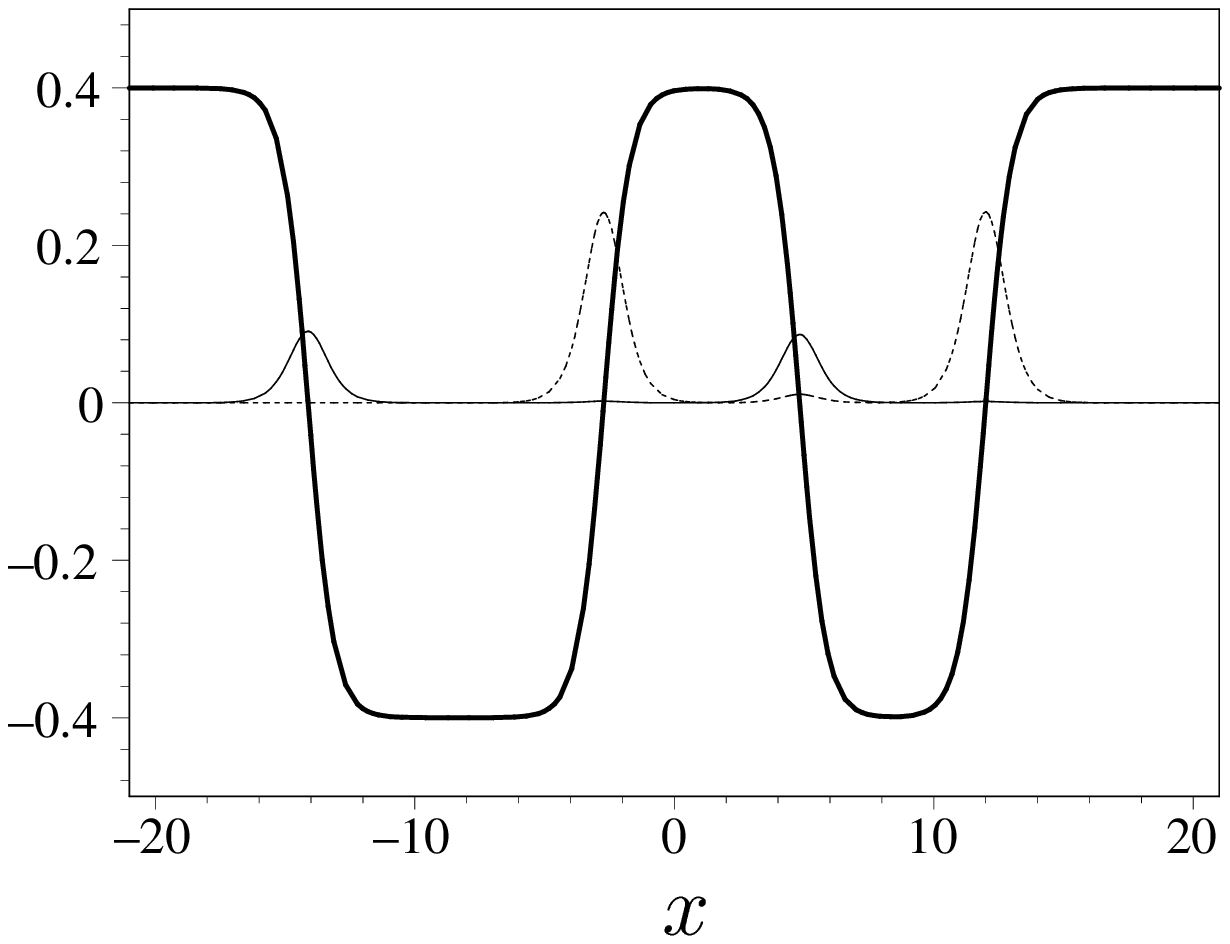,angle=270,width=5cm}
\epsfig{file=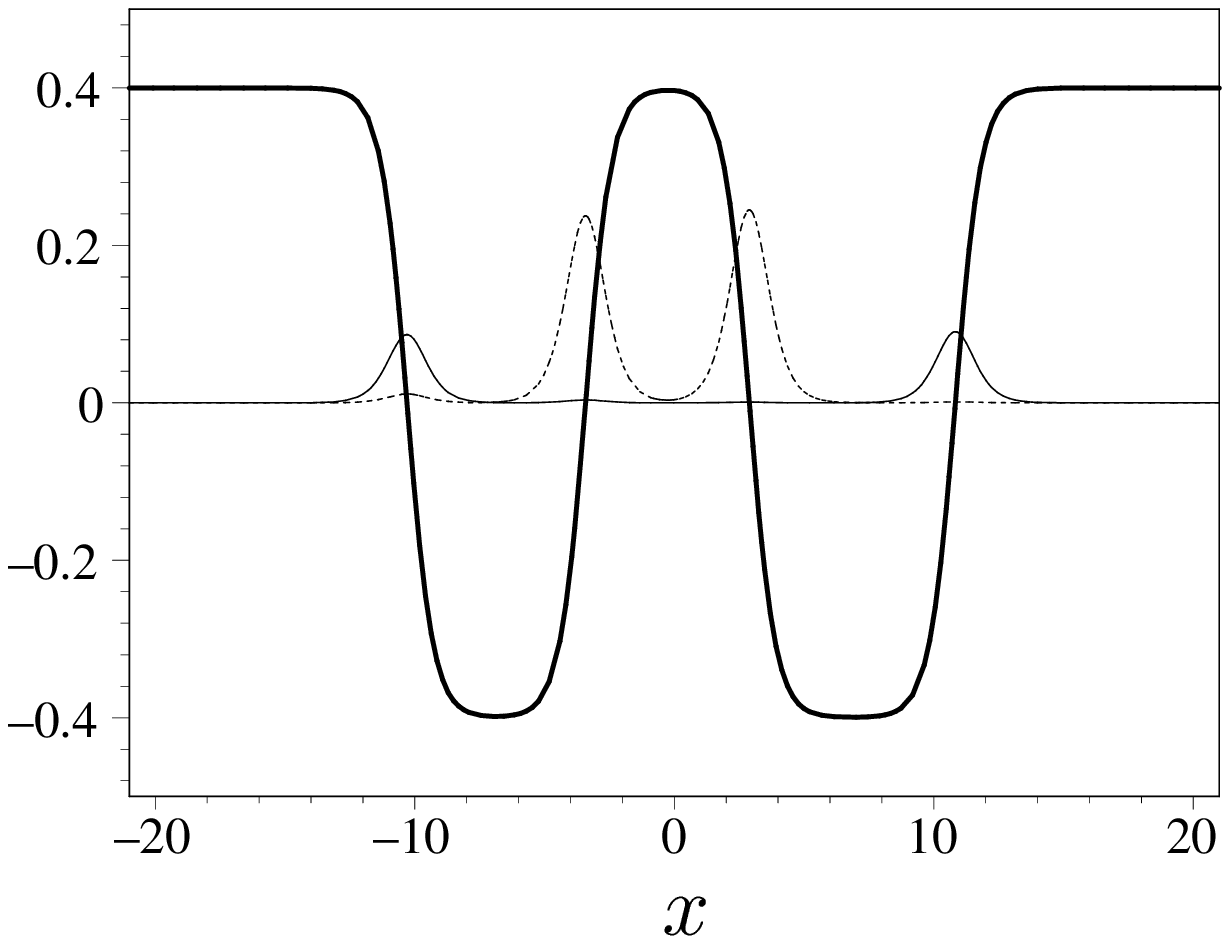,angle=270,width=5cm}\\
\epsfig{file=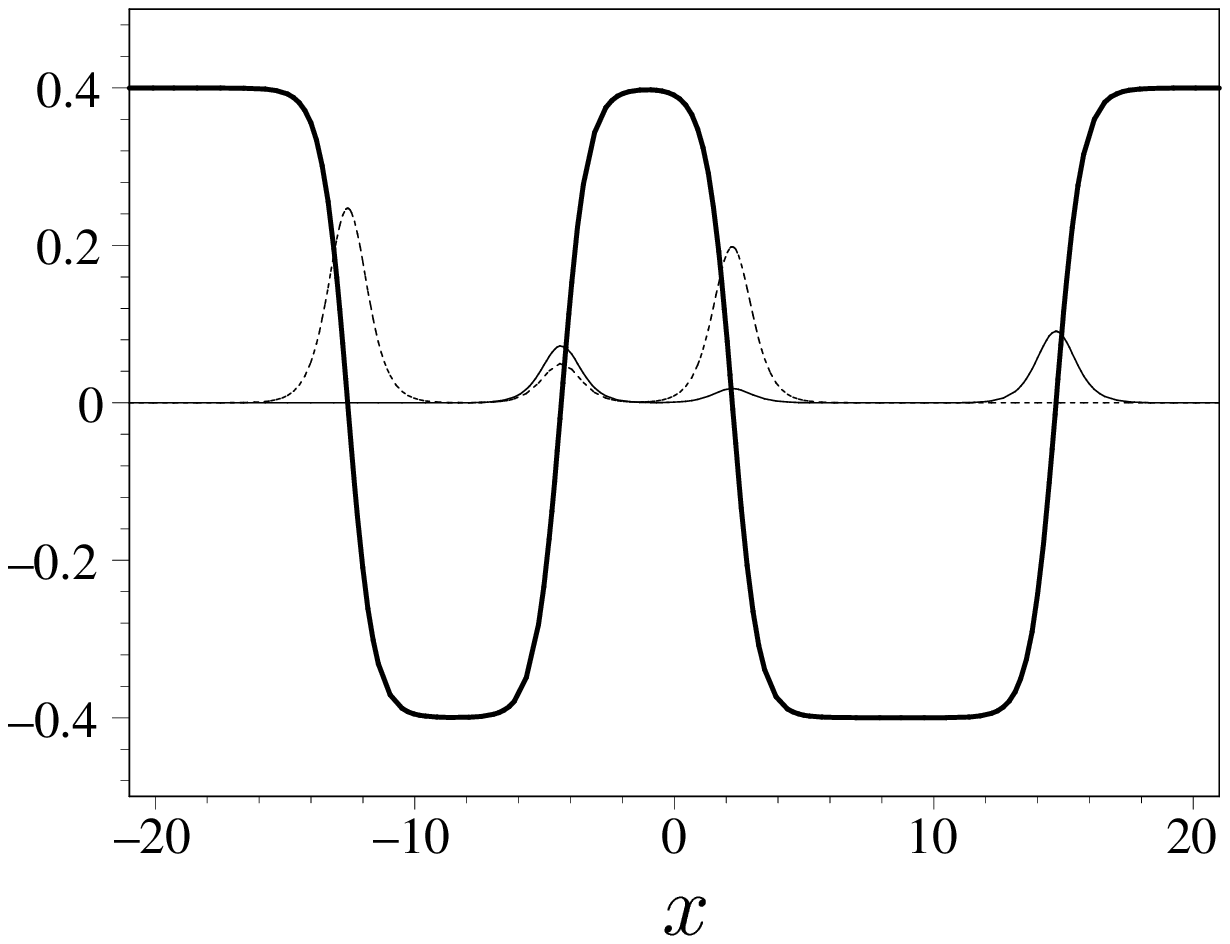,angle=270,width=5cm}\epsfig{file=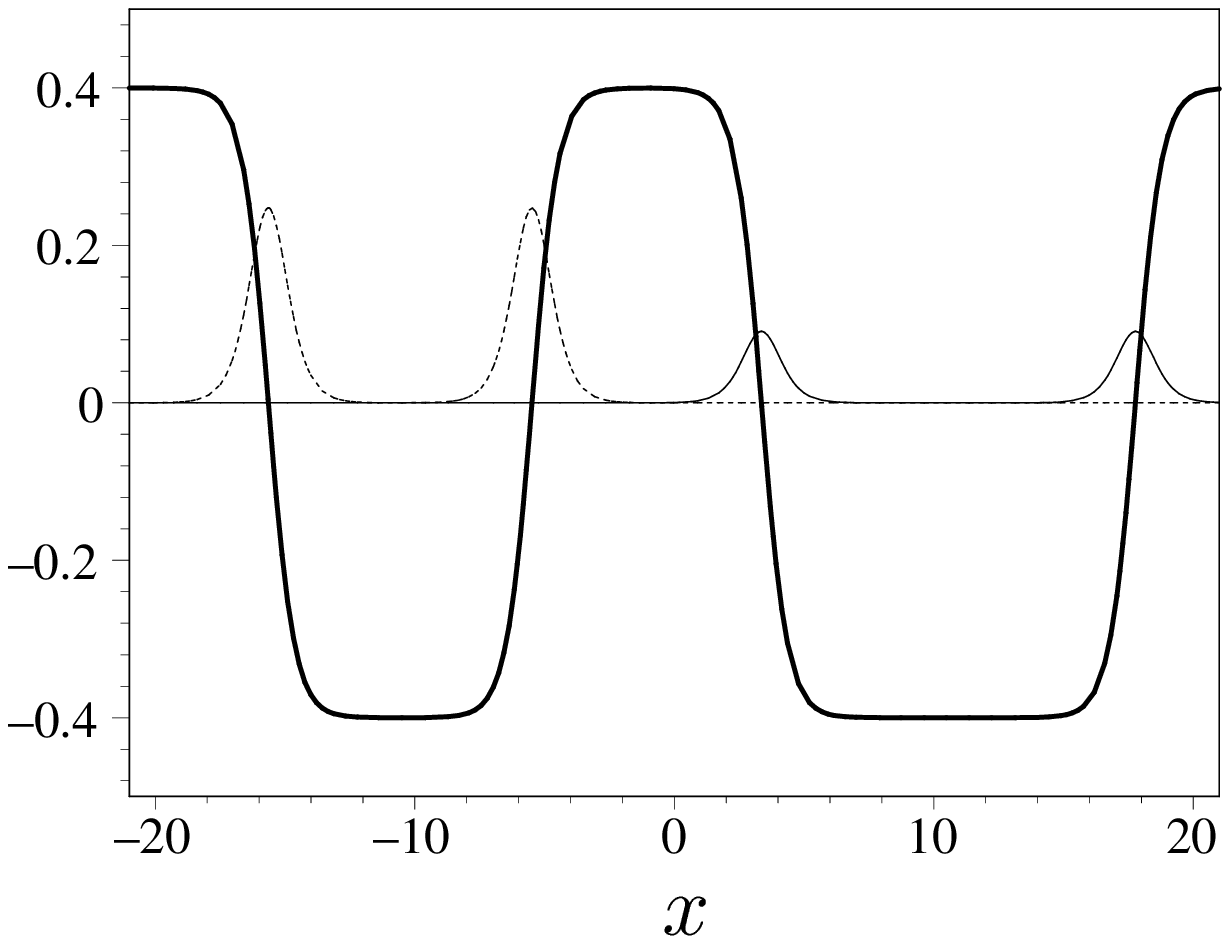,angle=270,width=5cm}
\caption{Snapshots of the collision process of Figs.~12 and 13 at times -2000,-1500,-1000,-500,0. Fat line: $0.4\times S$, solid thin line: $\rho^{(1)}$,
dashed thin line: $\rho^{(3)}$. See main text for a discussion of tunneling processes.}
\label{Fig14}
\end{center}
\end{figure}

\section{Summary and outlook}\label{sect4}
In this paper, we have extended previous works on baryon-baryon scattering to breather-breather scattering in the massless GN model. We have first
introduced the breather originally due to DHN and illustrated its behavior in space and time with a few examples. It can be regarded as a vibrational excited
state of the baryon of semi-classical type, characteristic for the large $N$ limit. We then had to recast it into a form better suited for the scattering problem.
We used an ansatz method developed in the context of baryon scattering problems. The basis exponentials entering the joint ansatz for scalar potential
and spinors in the TDHF approach have to be complexified, otherwise everything goes through as before. In this way one can derive exact, analytical
expressions for breather-breather scattering with arbitrary initial conditions and parameters. This contains also breather-baryon or baryon-baryon scattering
as special cases, as well as bound state problems if the velocities of both scatterers are chosen to be equal. The calculation involves a large number of coefficients
which have to be determined partly asymptotically (from single breather input), partly by solving the Dirac equation algebraically. We have made every effort to 
present
the results in the most compact form. While doing this, we encountered a number of simplifications and observed more algebraic structure than anticipated. 
We do not yet fully understand these features, but they point to the possibility of simplifying further our calculation. Right now, somewhat ironically,
the computations with Maple are perhaps the lesser problem as compared to the task of presenting the results in a digestable form.

Since breathers are kind of exotic objects, at least in particle physics, one may ask whether it is worth the effort to study them in such great detail.
We think that if one is interested in solving the GN model as completely as possible, there is no way around considering breathers. In this context,
it is instructive to look at the simpler problem of the classical sine-Gordon equation for a moment. There, one may ask what is the most general multisoliton
solution
corresponding asymptotically to spacelike separated individual solitons (i.e., disregarding solutions with a finite density of solitons like soliton crystals).
This has been answered by inverse scattering theory some time ago. The result is a known algebraic solution consisting of any number of interacting solitons 
and breathers, all with different velocities \cite{L19}. If we ignore the breathers for a moment, this can be compared to the general multi-baryon solution of the
GN model
discussed in Ref.~\cite{L9}. The main difference is that in the GN model, two baryons can have the same velocity if they have different fermion numbers, thus
describing multikink-bound states absent in the sine-Gordon model. Since breathers also arise in the GN model, it is plausible that the most general TDHF 
solution of the GN model (with a finite number of solitons) will also consist of breathers and solitons. In the baryon problem,
we found that the only input needed to tackle the $n$-baryon problem are the solutions of the one- and two-baryon problems. This was interpreted as signature
of factorization on the level of the composite states. Likewise, one might expect that a similar generalization exists with breathers, and that the 
two-breather problem solved here is sufficient to deal with any number of breathers, using again factorization. If correct, this would indicate that the
two-breather scattering problem solved here may actually play a central role and take us a long way towards the most general TDHF solution. As a side benefit,
we would also be able to generate the most general transparent potential of the Dirac equation, a problem similar to the one which has been solved 
for the Schr\"odinger equation long ago by Kay and Moses \cite{L20}.

\vskip 1.0cm
We should like to acknowledge helpful discussions with Gerald Dunne. This work has been supported in part by
the DFG under grant TH 842/1-1.

\end{document}